\documentclass[12pt]{amsart}
\usepackage{fullpage}
\usepackage{float,rotating}
\usepackage{caption}
\usepackage{parskip}
\usepackage[dvipsnames]{xcolor}
\usepackage{subfig}
\usepackage{breqn}
\newenvironment{tablenotes}[1][Note]{\begin{minipage}[t]{\linewidth}\footnotesize{\itshape#1: }}{\end{minipage}}
\PassOptionsToPackage{hyphens}{url}
\usepackage[
colorlinks = true,
linkcolor = blue,
urlcolor  = blue,
citecolor = blue,
anchorcolor = blue
]{hyperref}

\usepackage{graphicx}
\usepackage{longtable}
\usepackage{booktabs,array,multirow}

\usepackage[noamsmath]{kpfonts}
\usepackage{inconsolata}
\usepackage[T1]{fontenc}
\usepackage{amsfonts,amsmath,amssymb}
\usepackage{amsbsy}
\usepackage{amstext}
\usepackage{amsthm}
\usepackage{nicefrac}
\usepackage{enumitem}

\newtheorem*{corrhypothesis}{Correlated-Basin/Null-effect Hypothesis}
\newtheorem*{indhypothesis}{Independent-Basin Hypothesis}

\newtheorem{result}{Result}

\newif\iflatexml\latexmlfalse
\AtBeginDocument{\DeclareGraphicsExtensions{.pdf,.PDF,.eps,.EPS,.png,.PNG,.tif,.TIF,.jpg,.JPG,.jpeg,.JPEG}}
\usepackage{url}

\usepackage[authoryear,longnamesfirst]{natbib}
\usepackage{array}

\newcolumntype{P}[1]{>{\raggedleft\arraybackslash}p{#1}}
\newcolumntype{C}[1]{>{\centering\arraybackslash}p{#1}}
\newcommand{\SD}[1]{{\tiny $\left(#1\right)$}}

\newcommand{\mEst}[2]{\underset{\mbox{\SD{#2}}}{#1}}
\newcommand{\Prob}[1]{\mbox{Pr}\left\{#1\right\}}

\newcommand{\EStrat}[1]{\alpha_{\text{\tiny #1}}}

\newcommand{\NiceSetBig}[1]{\Bigl\{ #1 \Bigr\}}
\newcommand{\minitab}[2][l]{\begin{tabular}{#1}#2\end{tabular}}

\newcommand{\Treat}[2]{$\left({ }^{#1}_{#2}\right)$}
\newcommand{\DeltaP}[1]{\Delta p^{\star}_{\text{\footnotesize{#1}}} }

\usepackage{setspace}

\def\payMatrix{0}%0 Flag shows both in output. Otherwise 1 shows Hard, 2 shows Easy
\def\groupSize{0}%0 Flag shows both in output. Otherwise 2 shows 2 player, 4 shows 4 player, 10 shows 10 player
\def\chatTreat{0}%3 Flag shows both in output. Otherwise 0 shows non Chat, 1 shows Chat 
\def\terminalTreat{2}%0 Flag shows both in output. Otherwise 1 shows 50, 2 shows 75
\def\antibTreat{1}%0 Flag shows both in output. Otherwise 1 shows Bertrand, 2 shows anit-Bertrand
\newcommand{\groupText}[3]{
\if\groupSize0{\{\textcolor{red}{#1}\}\{\textcolor{blue}{#2}\}}
\else\if\groupSize2{#1}
\else\if\groupSize4{#2}
\else{#3}\fi\fi\fi}
\newcommand{\payText}[2]{\if\payMatrix0{\{\textcolor{ForestGreen}{#1}\}\{\textcolor{BurntOrange}{#2}\}}\else\if\payMatrix1{#1}\else{#2}\fi\fi}
\newcommand{\chatText}[2]{\if\chatTreat3{\{\textcolor{Aquamarine}{#1}\}\{\textcolor{Plum}{#2}\}}\else\if\chatTreat0{#1}\else{#2}\fi\fi}
\newcommand{\antibText}[2]{\if\antibTreat0{\{\textcolor{Aquamarine}{#1}\}\{\textcolor{Plum}{#2}\}}\else\if\antibTreat1{#1}\else{#2}\fi\fi}
\newcommand{\terminalText}[2]{
\if\terminalTreat0{\{\textcolor{Aquamarine}{#1}\}\{\textcolor{Plum}{#2}\}}
\else\if\terminalTreat1{#1}
\else{#2}\fi\fi}
\begin{document}
%  Extending and Validating Theoretical measures for Equilibrium selection in Dynamic Settings: Incorporating $N$
\title{Testing Models of Strategic Uncertainty: \\
Equilibrium Selection in Repeated Games}
%\title{Extending and Testing Equilibrium-Selection Models for Repeated Games: Dimensions of Substitution for Strategic Uncertainty}
% \title{Understanding Equilibrium Selection in Repeated Games: Strategic Uncertainty Substitutes}
% CUTE TITLES
%\title{Too many Cooks Spoil the Collusion? Examining the Effects of Strategic Uncertainty in Repeated Games}
%\title{Four's a Crowd}

% BORING TITLES

\author{Emanuel Vespa}
\author{Taylor Weidman}
\author{Alistair Wilson}
\thanks{We would like to thank: David Cooper, Guillaume Fr\'echette, Daniella Puzzello and Lise Vesterlund. This research was funded with support from the National Science Foundation (SES:1629193).}
\date{January, 2021}
\begin{abstract}
In repeated-game applications where both the collusive and non-collusive outcomes can be supported as equilibria, researchers must resolve underlying selection questions if theory will be used to understand counterfactual policies. One guide to selection, based on clear theoretical underpinnings, has shown promise in predicting when collusive outcomes will emerge in controlled repeated-game experiments. In this paper we both expand upon and experimentally test this model of selection, and its underlying mechanism: strategic uncertainty. Adding an additional source of strategic uncertainty (the number of players) to the more-standard payoff sources, we stress test the model. Our results affirm the model as a tool for predicting when tacit collusion is likely/unlikely to be successful. Extending the analysis, we corroborate the mechanism of the model. When we remove strategic uncertainty through an explicit coordination device, the model no longer predicts the selected equilibrium.
\end{abstract}

\maketitle
%; as well as seminar audiences at Indiana, Essex and the Behavioral Game Theory Conference
\section{Introduction}
Assumptions about which of the possible equilibria best captures the participants' economic behavior take on a central role in applications with repeated interaction. For example, in models of oligopoly where firms interact repeatedly both collusive and non-collusive equilibria can arise. Researchers must often find evidence in the data to identify which equilibrium best captures the status quo. But  equilibrium-selection problems persist when trying to predict counterfactual behavior, where appeals to the data are not possible. Consider a market where the researcher can establish that the current behavior of firms matches the non-cooperative equilibrium, but where their goal is to predict the effect from industry consolidation on consumer welfare. Embedded within the question is a concern on whether the selected equilibrium remains the same or will move under the counterfactual. Were a change isolated solely to the number of firms in the market, a directional comparative-static on selection could be posited: With fewer firms in the market, ceteris paribus, the equilibrium is likely to be pushed toward greater collusion.

But counterfactual analyses are often more involved than a simple comparative static. In our oligopoly example, while a lower number of firms might push behavior in the collusive direction, consolidation tends to be accompanied by other concurrent changes. For example, bigger firms can generate efficiency gains, increasing the marginal returns to price competition and product innovation. These accompanying shifts to the profit functions could plausibly pull more towards the non-collusive equilibria.\footnote{The industry-consolidation literature provides ample evidence that mergers are regularly accompanied by other changes in addition to a lower number of market participants. For instance, accompanying effects include lower costs  \citep{farrell1990horizontal}, higher capacities \citep{perry1985oligopoly}, new product offerings \citep{deneckere1985incentives}, and asymmetry among firms \citep{compte2002capacity}.} As such, consolidation has the potential for conflicting effects on the likelihood of collusion. More broadly, whenever counterfactual policies involve richer changes to more than one variable, predicting the resulting effect on the selected equilibrium can become substantially more challenging.

In this paper we evaluate a model of equilibrium selection constructed over the theoretical primitives; a model capable of integrating counterfactual shifts to many primitives at the same time. In doing this, we build on experimental studies of the infinitely repeated prisoner's dilemma (RPD),  where strategic uncertainty has been offered as an explanatory model  for rationalizing several findings across this literature. Specifically, \citet{DalBoFrechette2011} outline the organizational power of a selection notion from \citet{harsanyi1988general} when applied to data from their experimental treatments.\footnote{A subsequent meta-study across the literature \citep{dal2018determinants} confirmed the pattern.} The central idea here is to formally model the strategic uncertainty surrounding the collusive, history-dependent, equilibrium. In so doing, the analyst can reduce the equilibrium-selection problem to an understanding of the risk/reward trade-off over collusion attempts. The developed strategic-uncertainty measure is therefore a pure function of the primitives of the game. Thus, the model can help understand the trade-offs between two or more competing shifts of the primitives---for example, a change to the payoffs that increases the rewards to successful collusion combined with a reduction to the discount rate that makes such an attempt riskier. The model mediates the effects of any multivariate change into a single dimension, the effect on strategic uncertainty, which can then be used to predict the eventual likelihood of collusion. However, while the literature has shown the explanatory power of the strategic-uncertainty measure, previous studies have not been designed to test it, to examine the mechanism, and its potential limitations. 

To stress test this equilibrium-selection model we expand the set of primitives under consideration. In so doing we introduce a new source of strategic uncertainty to the repeated environment, one that plays an important role in applications, as motivated by our industry consolidation example above. Similar to the original two-player RPD experiments, our treatments will manipulate payoff primitives (primarily a stage-game payoff-parameter $x$) that affects the riskiness of collusion. But we also manipulate the number of players within the environment ($N$), where collusion in the model becomes riskier in larger groups. Hence, we can study joint changes in both $x$ and $N$---changes that push the equilibrium selection in two different directions, where the model's ability to predict the joint effect is the object of interest. As such, we will test the main hypothesis that equilibrium-selection across two distinct sources can be understood through the strategic-uncertainty model. We therefore construct treatment manipulations that modify $x$ and $N$ relative to a baseline in a way that the strategic-uncertainty measure is held exactly constant. Finding constant cooperation across the double-barreled change would be consistent with the modifications to $x$ and $N$ being perfect substitutes, per their modeled effect on strategic uncertainty.\footnote{Our treatments also include manipulations where the strategic-uncertainty effects of one of the two variable changes theoretically dominates the competing effect, with a predicted effect on selection. Moreover, the design also embeds comparative-static comparisons where a single variable is manipulated.} Complementing the construction and test of the core strategic-uncertainty model, our design also embeds an alternative hypothesis (which we characterize the behavioral underpinnings of) in which the group-size variable $N$ does not affect strategic uncertainty, with a clearly distinct set of predictions from our main model.

Our main results allow us to clearly separate and validate the equilibrium-selection model based on strategic uncertainty. Across the six possible treatment comparative-statics in our design, four are predicted to have clear directional effects on selection (with the other two being perfect substitutes). In every single one of these four comparisons, data from our experiments move in the predicted direction, with effects sizes that are both statistically and economically significant. This is true for both initial and ongoing behavior for the repeated games we examine. The more-challenging comparisons involve treatments with compound shifts that move selection in opposed directions. Two of our treatment comparisons are constructed so that changes across the payoff and group-size are perfectly substituting as captured by the strategic-uncertainty model. Relative to successful coordination on the collusive outcome, the metric that likely matters most for applications, our results show strong support for the selection model, and our designed shifts to the payoffs and number of players are direct substitutes. Meanwhile, we document discrepancies between the measure's predictions and behavior in the first period of interaction. As such, our results indicate that the extended selection model is more accurate in predicting ongoing collusion than in predicting initial intentions---where changes in $N$ provide the key variation to identify this. Overall, our results validate the theoretical model, providing a tool to understand which effects might dominate in counterfactuals where policy shifts over multiple primitives. In particular, the results lend credence to the idea that a reduction of the environment to a single dimension capturing strategic uncertainty can be predictive of equilibrium selection.

Uncertainty regarding the other players' actions is put front-and-center in our main treatments as the driver of equilibrium selection. Hence, if strategic uncertainty is the causal mechanism driving selection, then removing or reducing the doubts about others' play should make the selection model's predictions moot. We pursue this idea in an additional set of treatments where we allow for pre-play communication in a setting where we would otherwise expect little collusion. Participants are given the opportunity to exchange free-form messages prior to the repeated game, a feature that can be used to reduce uncertainty over others' intentions \citep[see][]{kartal2018new}. For the same experimental parameterization where we observe ongoing cooperation rates below one percent without communication, the effect from adding it is to shift behavior to the other extreme: with initial (ongoing) cooperation rates of  95 (80) percent.\footnote{As a check that the provided communication channel is not driving results separate from the equilibrium coordination effects, we implement a second treatment with communication but where the collusive outcome is not a robust equilibrium outcome. Here we find that communication does not lead to successful collusion. In other words, communication only has an effect when there are clear theoretical incentives for collusion.} Given these results, we conclude that strategic uncertainty, the mechanism underlying the model, is the causal channel. In addition to mechanism validation, these results clearly outline the limitation of the model to understanding \emph{tacit} collusion, in the sense that it fails to provide useful guidance when collusion is more explicit.

In terms of the paper's organization, we next outline the related literature. In section \ref{sec:theory} we outline our generalization of the selection model from the two-player repeated PD game, outline the predictions from that literature, and provide design details. In section \ref{sec:results} we analyze the results from our core treatments and demonstrate the model's predictive properties generalize to $N$. Section \ref{sec:extensions} then examines some extensions: one showing that our results hold for both between- and within-subject identification; and the other pinning down the strategic uncertainty mechanism via the opportunity for explicit collusion. Finally, in section \ref{sec:conclusion} we conclude.

\subsection{Literature}
Our paper builds on the recent consolidation of the experimental RPD literature presented in \cite{dal2018determinants}. While one of our baseline treatments replicates a standard finding in the literature,\footnote{As pointed out in \citet*{berry2017assessing} many experimental replications become harder to find where the papers do not explicitly point out the replication components.} our core contribution is to generalize the equilibrium selection model outlined in the meta-study, adding an additional source of strategic uncertainty: the number of players, $N$.\footnote{The extension of the notion of equilibrium selection described in \citet{harsanyi1988general} to the RPD was first proposed by \citet{blonski2015prisoners} and first shown to organize data by \citet{DalBoFrechette2011}. See also \citet{fudenberg2010slow} for an examination of the effects with imperfect monitoring, and \citet{kartal2018new} for a test of a selection theory based on individual heterogeneity in preferences over dynamic strategies.} Where the literature has developed this model to be explanatory, our approach is both to expand the selection model to a new setting, but also to take the model as the core experimental object to be tested. 

Our generalization of the strategic uncertainty model in the RPD literature is carried out in two alternative ways. The first extension (and most-standard, given its use of independent beliefs) formalizes a distinct source of strategic uncertainty from the payoff-based source in the meta-study. An alternative extension (based on fully correlated beliefs) reflects a null effect, that the newly introduced source has no effect. As such, our generalization offers a potentially profitable design approach for future research examining other channels for strategic uncertainty effects---asymmetries in the action space or payoffs, the effects of sequentiality, etc.\footnote{See \citet{ghidoni2020effect} and \citet{kartal2018new} for experimental examinations the effect sequentiality has in RPD settings through a reduction in strategic uncertainty.} 

Our environment also allows us to better distinguish between the empirical measures linked to the selection model. That is, using literature-level data assembled by \citet{dal2018determinants}, we show that the two-player RPD strategic uncertainty model predicts both initial and ongoing cooperation well. However, neither is identified particularly well from the other. With more than two players though, this is no longer the case, where we use this to demonstrate that the developed strategic uncertainty model is better suited for predicting successful collusion rather than initial intentions.\footnote{Ongoing cooperation is a measure that is likely to be more relevant for empirical applications where collusion may be a worry. For instance, from \citet{harrington2016relative}, page 256: ``(...) collusion is more than high prices, it is a mutual understanding among firms to coordinate their behavior. (...) Firms may periodically raise price in order to attempt to coordinate a move to a collusive equilibrium, but never succeed in doing so; high average prices are then the product of failed attempts to collude.}

This paper is part of a broader literature seeking to understand and document regularities in equilibrium selection. In particular, regularities that are amenable to theoretical modeling. To this end, the strategic-uncertainty measure that we examine is particularly promising, as the equilibrium objects required for calculation are computationally simple: the stationary non-collusive equilibrium and the history-dependent collusive equilibrium. In environments beyond the RPD where the equilibrium outcomes are held constant, the model can be similarly extended per our illustration with a move to $N$ players. However, in more complex environments where the equilibrium set changes more substantively, the constraint to two focal equilibria in the strategic uncertainty model may lose validity, and/or raise questions as to which two strategies are focal. Examples of more-complex settings includes dynamic games, where the stage environment can change across the supergame and the space of strategies becomes substantially larger. \cite{vespa2020experimenting} focus on a horse-race examination of which two equilibria are focal (from a wider set of possible alternatives) for rationalizing behavior in dynamic games. The paper identifies a similar strategic uncertainty measure constructed around the most-efficient Markov perfect equilibrium and the best \emph{symmetric} collusive equilibrium. A strategic-uncertainty model based on these strategies predicts behavior, where these strategies dovetail with the repeated-game strategies in the simpler environment studied here.\footnote{Work on equilibrium selection in dynamic games builds on recent contributions in this area. For example: \cite{battaglini2011legislative,battaglini2016dynamic,agranov2016static,kloosterman2019cooperation,vespa2019experimenting,rosokha2020cooperation,salz2020estimating,vespa2020experimental}.} 

An experimental literature on behavior in oligopolies documents that collusion clearly responds to the number of players. Both Cournot \citep{huck2004two,horstmann2018number} and Bertrand settings \citep{dufwenberg2000price} indicate that as the number of players increases collusion becomes less likely, often as soon as $N$ exceeds two.\footnote{See also references in \cite{potters2013oligopoly} for similar findings.} We contribute to this literature on two margins. First, papers here focus on out-of-equilibrium behavior in settings with a finite-time horizon, and a subsequently unique theoretical prediction. In contrast, we examine how changes to $N$ affect outcomes in an infinite-horizon environment with both collusive and non-collusive equilibria. Second, and crucially, our focus is not just on the qualitative directional effects from $N$, but on validating a model of how it affects strategic uncertainty. The model allows for an understanding of the extent of substitutability between primitives under consideration; and where validated can predict the directional effects of more-nuanced, multi-dimensional counterfactuals. Clearly, any equilibrium-selection notion suggested for such a task requires a great-deal of scrutiny. However, our findings do suggest substantial optimism that this path may be fruitful.

Our work is also related to an experimental literature on mergers that manipulates the number of players. As surveyed by \citet{goette2009merger}, some experiments deal with a ``pseudo-mergers,'' where a subset of the original firms remain in the market \citep[for example]{huck2007merger}. Others experiments implement ``real mergers,'' where mergers  introduce other changes in the market beyond $N$ \citep{davis2002strategic}. A contribution of our approach is that the strategic-uncertainty measure can predict counterfactual behavior in both settings. Another discussion in this literature is whether merger effects are evaluated within the same group of participants (within-subject designs) or across different groups (between-subject design). In this paper, while our main treatments make use of between-subject identification, we also conduct within-subject sessions at the same parameterization, demonstrating that though there can be meaningful short-run differences, with enough experience the results do not depend on the approach.\footnote{Differences in behavior can be stickier if changes are small or introduced gradually. \cite{weber2006managing} shows that increasing the number of players gradually in a coordination game leads to different results relative to a situation when play starts with a large group. In repeated games gradual introduction of changes to the payoff primitives have also been found to have effects in repeated games, see \cite{kartal2017building}. These results suggest that the selection-notions we are examining are relevant for `large' counterfactual changes, and where future research can help clarify how to integrate `large' into a predictive model of selection.}

The effects of communication devices as a bolster for collusion are well-established in the experimental literature. As surveyed in \cite{cason2008price} and \cite*{harrington2013relative}, more-structured, limited forms of communication usually result in small, temporary collusive gains, where free-form communication generates large, long-lasting effects.\footnote{For further details on the effect of communication in repeated games with an unknown time horizon, see \cite{fonseca2012explicit}, \cite{cooper2014communication}, \cite{harrington2016relative}, \cite{wilson2020information} and \cite{frechette2020multimarket}.} For these reasons, in one of our extensions we examine unrestricted chat messages as a strong coordination device. The collusive results here clearly indicate that the domain for our strategic-uncertainty measure based on tacit-collusion does not include environments where explicit collusion is likely. However, we do show that there are clear limits on the effects of explicit collusion, and that these limits are directly predicted by theory. Using a change to the payoff primitives (here the discount rate) we make collusion a knife-edge, non-robust equilibrium, and show that the effects of communication are entirely dissipated.

While much of the literature on repeated games studies the standard two-player PD game, there is a large literature studying a canonical $N$-player social dilemma: the voluntary contribution public-goods game \citep[see][for a survey]{vesterlund2016using}. Though much of this literature focuses on finite implementations, one notable exception is \citet{lugovskyy2017experimental}. Similar to our paper, the authors use experimental variation over both $N$ and the payoff primitives (in their case the return to group contribution). However, this is done with a different end goal. They design offsetting variation in the payoffs and $N$ to identify the isolated effect of the stage-game's MPCR. We instead do so to isolate strategic uncertainty, to test a predictive theory of selection.

Beyond social dilemmas, our paper is also connected to the literature on coordination games  \citep[see][for a survey]{devetag2007and}. The strategic-uncertainty measure examined in our paper works because the infinitely repeated prisoner's dilemma game has a stag-hunt normal-form representation \citep{blonski2015prisoners}, adapting the risk-dominance notion for one-shot coordination games in \citet{harsanyi1988general}.\footnote{The difference in our setting is that neither total payoffs nor strategic choices are directly provided to participants, as these are extensive-form objects. Instead they are provided with the stage-game payoffs/actions, where strategies (such as the grim trigger or tit-for-tat) and gross payoffs must be endogenously formulated.} Risk dominance (and the cardinal implementation through the measure of strategic uncertainty) has been shown to have substantial predictive content in simple coordination games with trade-offs over payoff-dominance and risk-dominance \citep[see][and references therein]{battalio2001optimization,brandts2006change}. Strategic uncertainty has therefore demonstrated its usefulness as a theoretical selection device both in directly presented one-shot games, and in repeated games. Our contribution to this literature is to design an experiment to explicitly test it and show that the predictive effects extend further still, to multi-player infinite-horizon settings.

\section{Generalizing the Basin of Attraction\label{sec:theory}}
Developing empirical criteria for equilibrium selection in games where collusion is possible requires two measures: one theoretical, one empirical. On the theory side we need a prediction, a model that maps the primitives of the game into a scalar where upward/downward movements are clearly interpretable as increasing/decreasing the likelihood of collusion. On the empirical side we need a precise target against which the theoretical notion can be contrasted and validated. This empirical measure should examine a behavior that differs starkly under the collusive and non-collusive equilibria. 

We begin this section by summarizing the progress made towards validating a theoretical prediction in the two-player infinitely repeated prisoner's dilemma (RPD) literature. The theoretical notion here is the size of the basin of attraction for the strategy \emph{always defect}. The focal outcome measure is the cooperation rate of individual players in the initial rounds. In what follows we discuss the issues as we extend the exercise to a new source of strategic uncertainty, the number of players $N$---both for the theoretical and empirical measures.

\subsection{Two-players}
Consider an RPD with discount rate $\delta\in(0,1)$. In each period $t=1,2,\ldots$ players $i\in\left\{1,2\right\}$ simultaneously select actions $a_i\in \mathcal{A}:=$\{(C)ooperate, (D)efect\}. The period-payoff for player $i$ is a function of both players' choices, $\pi_i(a_i,a_j)$, where  all symmetric PD games can be expressed in a compact form by normalizing all payoffs relative to the joint-defection payoff $\pi_0=\pi(D,D)$, and rescaling with the relative gain from joint cooperation: $\Delta\pi:=\pi_i(C,C)-\pi_0$.\footnote{More exactly, game payoffs $\pi$ can be transformed via $\tilde{\pi}=(\pi_i-\pi_0)/\Delta\pi$ to measure all payoffs relative to joint defection in units of the optimization premium.} Defining scale and normalization in this way, the PD stage-game can be expressed with two parameters $g$ and $s$ for the different-action payoffs $\pi_i(D,C)=\pi_0+(1+g)\Delta\pi$ and $\pi_i(C,D)=\pi_0-s\cdot\Delta\pi$. The  parameters  $g>0$  and $s>0$ thereby capture the relative temptation- and sucker-payoff parameters, respectively.

The RPD payoffs can be used as primitive inputs into a risk/reward model of collusion based upon strategic uncertainty. Strategic uncertainty here is distilled into a decision between two focal extensive-form RPD strategies:
\renewcommand{\labelenumi}{(\roman{enumi})}
\begin{enumerate}
\item \textit{Always defect}, $\EStrat{All-D}$, which plays the stage-game Nash in all rounds (the unique MPE of the game).\footnote{A Markov strategy is history independent, removing any conditioning on past play. A Markov-perfect equilibrium (MPE) is a subgame-perfect equilibrium (SPE) in which agents use Markov strategies. In an RPD, if choices are forced to be history independent then there is a unique equilibrium: playing the stage-game Nash equilibrium in all periods.} 
\item  The \textit{Grim Trigger},  $\EStrat{Grim}$, a strategy that begins by cooperating but switches to always-defect after observing any defections in past play (the best-case collusive SPE).\footnote{The strategy here is `best case' in three senses: (i) it can support the best-case outcome; (ii) it uses the harshest possible punishment, and so can support collusion at smaller values of $\delta$ than any other strategy; and (iii) any realized miscoordination is minimal, being resolved in a single round.} 
\end{enumerate}
As functions of the observable history $h_t$, these two strategies are given by:
\begin{align*}
    \EStrat{Grim}(h_t) & =\begin{cases}
    C & \text{if }t=1 \text{ or } h_t=((C,C),(C,C),\ldots,(C,C)), \\
    D & \text{otherwise;}
    \end{cases}\\
     \EStrat{All-D}(h_t)& =D.
\end{align*}

Strategic uncertainty in the two player RPD is measured through the \emph{size of the basin of attraction for always defect}. The model considers the expected payoff to player $i$ when uncertainty on the other player $j$ is represented by a believed strategy mixture $p\cdot\EStrat{Grim}\oplus (1-p)\cdot\EStrat{All-D}$. The basin for always defect is defined as the set of beliefs $p$ for which player $i$ yields a higher expected payment from  $\EStrat{All-D}$ than $\EStrat{Grim}$. The always-defect belief basin is therefore the interval $[0,p^\star(g,s,\delta)]$, where the critical-point/interval-width is given by:\footnote{
In the case that the strategy $(\EStrat{Grim},\EStrat{Grim})$ is not an SPE of the repeated game, the basin size for always defects is defined as $p^\star(x_T,x_S,\delta)=1$.}
\begin{equation}\label{TwoPlayerBasin}
    p^\star(g,s,\delta)\equiv\dfrac{(1-\delta)\cdot s}{\delta -(1-\delta)\cdot\left(g-s\right)}.
\end{equation}

The basin-size $p^\star(g,s,\delta)$ has a clear interpretation for strategic uncertainty: for any belief $p>p^\star(g,s,\delta)$ on the other player using the collusive strategy $\EStrat{Grim}$, the player does strictly better choosing $\EStrat{Grim}$; for any belief $p<p^\star(g,s,\delta)$, they do strictly better by selecting the non-collusive strategy $\EStrat{All-D}$. As such, the smaller $p^\star$ is, the lower the strategic uncertainty surrounding collusion. Moreover, the cardinal basin-size measure directly implies the ordinal risk-dominance relationship between the two strategies. If $p^\star(g,s,\delta)<\nicefrac{1}{2}$ then the collusive strategy $\EStrat{Grim}$ risk-dominates $\EStrat{All-D}$, and vice versa. Henceforth, by `basin of $p^\star$' we mean the maximal belief on the other cooperating for which the strategy $\EStrat{All-D}$ is optimal. 

Equation \eqref{TwoPlayerBasin} represents an easy-to-derive theoretical relationship between the payoff primitives of the game (here $g$, $s$ and $\delta$) and a critical strategic belief on the other player's likelihood of collusion. The hypothesized relationship is monotone, where the higher is $p^\star$, the lower is the likelihood of cooperation, thereby allowing for unambiguous directional predictions for any counterfactual change in the primitives. The posited mechanism within this model is also clear-cut: strategic uncertainty introduces a risk/reward trade-off for collusion attempts, which can be solved using standard economic analysis. This has two benefits: First, we can test the underlying  strategic-uncertainty mechanism through other channels outside of the model, where we will do exactly that in an extension examining coordination devices. Second, the necessary assumptions for extending this model to analyze alternative strategic-uncertainty sources are straightforward.

\begin{figure}[tb]
    \centering
    \subfloat[Initial cooperation]{
    \includegraphics[width=0.49\textwidth]{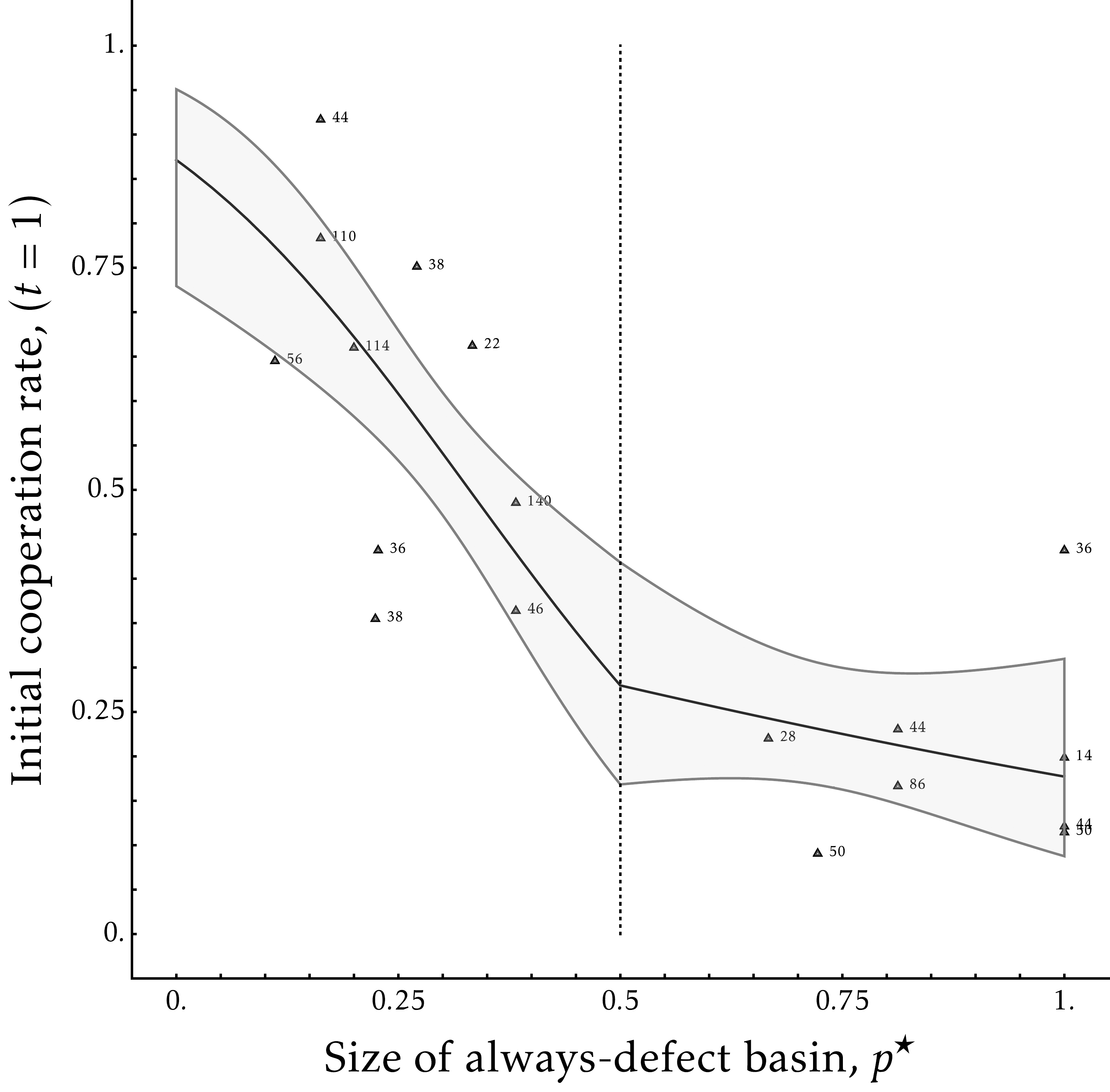}}
    \subfloat[Ongoing cooperation]{
    \includegraphics[width=0.49\textwidth]{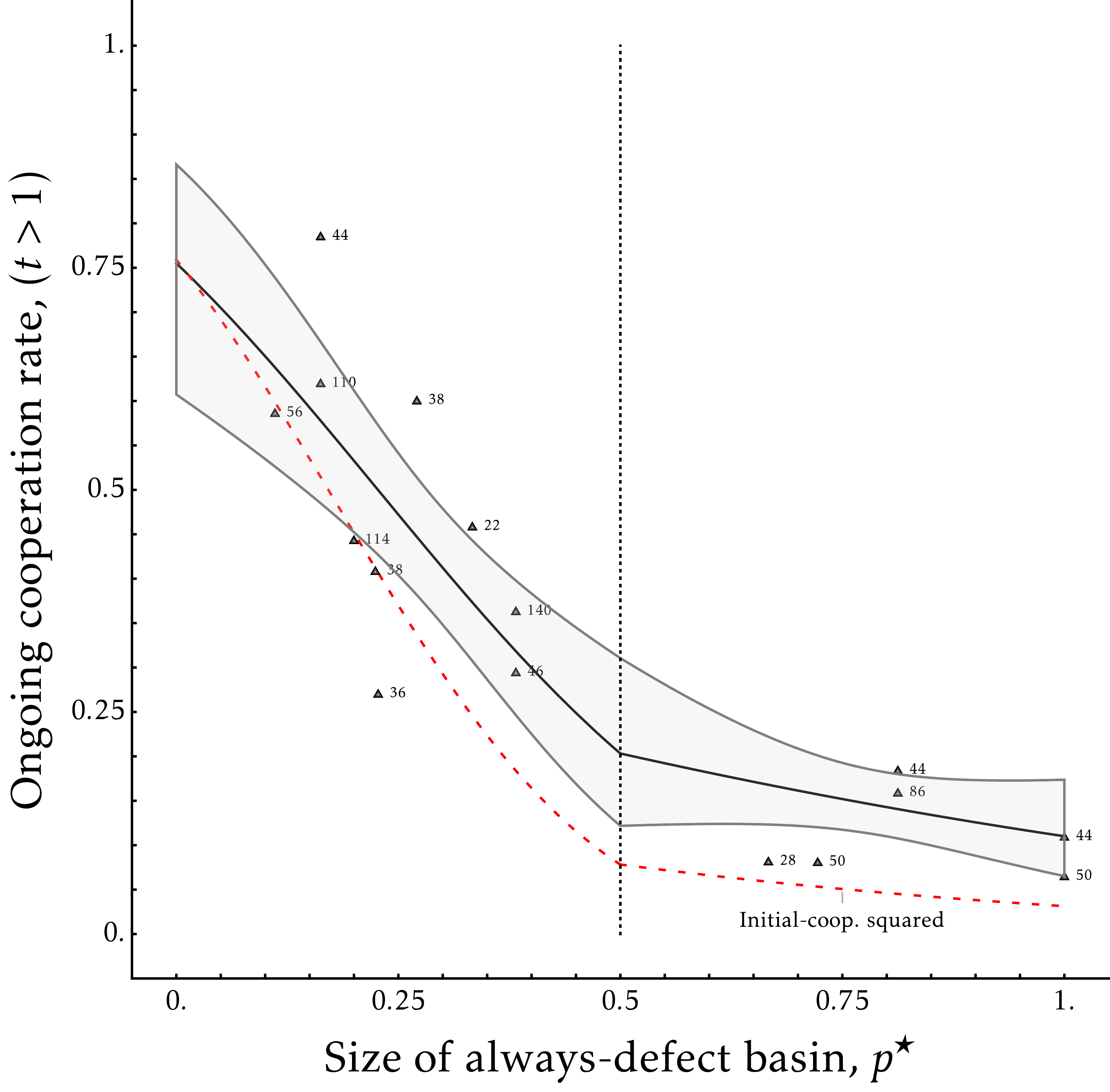}}
    \caption{Meta-study relationship: Strategic Uncertainty and RPD Cooperation}
    \label{fig:MetaStudy}
    
\begin{tablenotes}
Figures show estimated effects and 95-percent confidence intervals for initial/ongoing cooperation in RPD meta-study \citep{dal2018determinants}. Red dashed line in Panel (B) indicates implied theoretical relationship for ongoing cooperation based on the estimated relationship illustrated in Panel (A). Each point indicates a separate treatment, with the number of participants indicated.
\end{tablenotes}
\end{figure}

We now turn to the empirical criteria used to validate this theoretical measure. As a summary of the RPD literature we focus on the recent meta-study on the two player RPD \citep{dal2018determinants}. Two of the main results here are to show that the scalar basin-size measure of strategic uncertainty is highly predictive of behavior, though with a non-linear relationship. We illustrate this relationship in Figure \ref{fig:MetaStudy} for two empirical outcome measures. %In panel (A) we illustrate initial unilateral cooperation in the first round of the supergame (the measure the literature focuses on). In panel (B) we illustrate an alternative outcome, the ongoing unilateral cooperation in rounds two and beyond.  

In both panels in Figure \ref{fig:MetaStudy} the theoretical measure of strategic uncertainty (the basin-size $p^\star$) is presented on the horizontal axis. The empirical outcome measures are presented on the vertical axes. In Panel (A)  we present the results for \emph{initial cooperation} in the supergame,  a measure tracking collusive \emph{intentions} at the individual level as there is no interaction with matched partners at this point. In contrast, in Panel (B) we present results for \emph{ongoing cooperation}, choices in rounds two and beyond, a measure of the extent to which collusion attempts are successful. While there is no mechanical relationship between initial and ongoing cooperation, the theoretical model used in the basin construction posits an indirect linkage between the two. If collusion functions through conditional-cooperation with grim-trigger punishments, then the expected ongoing cooperation rate is the probability that the players jointly cooperate in the first round: the initial cooperation rate all squared. Thus, the posited relationship between initial and ongoing cooperation in the two-player RPD is not separately identifiable through any of the underlying payoff primitives.

Using the meta-study data from 996 participants across 18 experimental treatments, we estimate the relationship between late-session cooperation (supergames 16--20, the data we focus on in our experiments) and the strategic-uncertainty basin-size measure. Per the meta-study, the fitted relationship is estimated using a piecewise--linear probit model.\footnote{Individual-level cooperation decisions are the left-hand side variable, where the basin-size enters the right-hand side in a piecewise-linear fashion around the risk-dominance dividing point. The econometric specification is motivated by \citet[][Table 4]{dal2018determinants}; however to enforce level-continuity in the estimated relationship we remove a degree of freedom from their specification that allowed for a discontinuity at $p^\star=\nicefrac{1}{2}$.} The solid line in each figure indicates the predicted cooperation rate at each $p^\star$ from the probit estimates, where the shaded region indicates the 95 percent confidence interval for the prediction (clustering by treatment). For both initial and ongoing cooperation the same pattern is found: a consistently low cooperation level when always-defect is risk dominant ($p^\star>\tfrac{1}{2}$); and a significantly decreasing relationship with $p^\star$ when collusion is risk dominant ($p^\star<\tfrac{1}{2}$). As such, Figure \ref{fig:MetaStudy}(A) illustrates Results 3 and 4 from the meta-study.

Figure \ref{fig:MetaStudy}(B) indicates a similar qualitative relationship between strategic uncertainty and ongoing cooperation. Going beyond just the form though, layered on top of the estimated ongoing-cooperation relationship we also indicate the implied theoretical relationship with the initial-cooperation results. Using the square of the estimated \emph{initial} cooperation rate given in Panel (A) we therefore calculate the predicted \emph{joint} cooperation rate in round one. Under grim-trigger coordination, the initial joint-cooperation rate is exactly the ongoing cooperation rate, and so we illustrate this implied relationship in Panel (B) as the red dashed line. There are certainly some differences between the estimated relationship using actual ongoing cooperation decisions (black solid line) and the implied rate from initial cooperation (red dashed line). However, as the figure illustrates, these differences are mostly in levels for the region where always-defect is risk dominant.\footnote{In particular, we find significantly more ongoing cooperation when the basin measure is above $\nicefrac{1}{2}$ than we might expect from the initial rates and the use of a grim-trigger. This is consistent with another finding in the literature, the frequent use of more-forgiving/lenient strategies than grim. Tit-for-tat, for instance, can generate substantial \emph{ongoing} cooperation when matched with its variant that starts out defecting.} Slopes and levels for $p^\star$ in the $[0,\tfrac{1}{2}]$ region where the strategic uncertainty input has a clearer empirical response are not significantly different. The implication from this is that with only two players we cannot separately identify the extent to which the strategic-uncertainty measure is predicting initial intentions and/or successful coordination. However, as we will show, adding more players provides a channel to identify between the initial and ongoing measures.\footnote{For any setting where collusion requires $N$ agents to initially cooperate to produce ongoing cooperation then the relationship is simply initial cooperation rate to the $N$-th power. Separate identification between the two measures then comes about through comparison across treatments with different values of $N$.}

\subsection{Extending to \texorpdfstring{$N>2$}{N\textgreater 2}}
We now extend the strategic-uncertainty model above to an $N$-player environment. The core extension is intuitive: instead of considering one other player choosing a mixture $p\cdot\EStrat{Grim}\oplus (1-p)\cdot\EStrat{All-D}$, we consider $N-1$ other players choosing this same strategic mixture. 

To outline the extension, and set-up our design, we consider a family of symmetric social dilemmas that nest the standard RPD when $N=2$. However, to hold constant a $2\times2$ stage-game representation for all $N$, our family of dilemmas make use of an aggregate (and deterministic) signal of the other agents' actions. Each player $i=1,\ldots,N$ continues to make a binary action choice $a_i\in \mathcal{A}\equiv\left\{C,D\right\}$, but their payoffs do not vary with (and they do not receive feedback on) the separate actions of the other $N-1$ players. Instead, payoffs are determined by the own-action $a_i$ and a deterministic binary signal $\sigma(a_{-i})\in \left\{S\text{(uccess)},F\text{(ailure)} \right\}$ of the others' actions, $a_{-i}$. In particular, the generic player $i$'s stage-game payoff and signal function are given, respectively, by:
\begin{align*}
    \pi_i\Bigl(a_i,\sigma\Bigr) & =
    \begin{cases}
    \pi_0+\Delta\pi & \text{if  }a_i=C,\sigma=S,\\
    \pi_0+\Delta\pi\cdot\left(1+x\right) & \text{if  }a_i=D,\sigma=S,\\
    \pi_0-\Delta\pi\cdot x & \text{if  }a_i=C,\sigma=F,\\
    \pi_0 & \text{if  }a_i=D,\sigma=F;
    \end{cases}\\
    \sigma(a_{-i})& =
    \begin{cases}
    S & \text{if  }a_{j}=C \text{ for all }j\neq i,\\
    F & \text{otherwise.}
    \end{cases}
\end{align*}

These choices lead to a symmetric game, where payoffs can be summarized with a  $2\times2$ table over: (i) the own action $C$ or $D$; and (ii) the signal outcome, an $S$ signal if the other $N-1$ players jointly cooperate, or an $F$ signal if at least one matched player defects. The exact payoffs (with the same implicit scale $\Delta \pi$ and normalization $\pi_0$ as before) implement a PD-like environment where payoff-based strategic uncertainty is collapsed to a single parameter $x$.\footnote{In the meta-study notation this is implemented with $s=g=x$. This single-parameter formulation is equivalent to the \citet{fudenberg2010slow} benefit/cost formulation, where their benefit/cost ratio $b/c$ parameter is given by $\nicefrac{(1+x)}{x}$ here.} 

The success/failure signal here is a deterministic function of the $N-1$ other player's actions that dovetails with the standard RPD game when $N=2$. The choice for the signal function $\sigma(\cdot)$ here maximizes the coordinative pressure, duplicating a Bertrand-like tension: collusion is successful only when the other $N-1$ players all cooperate.\footnote{While potentially interesting, one disadvantage of a Cournot-like parameterization where the positive externality from others' cooperation is strictly increasing in the number of cooperators is that the cardinality of the signal/history spaces will then scale with $N$. However, note that the precise basin-size formula is \emph{identical} under the Cournot formulation if own cooperation has an additive cost $x\cdot\Delta\pi$ and the punishment path is triggered on \emph{any} level of defection. An alternative choice for a signal $\sigma(\cdot)$ that does not scale with $N$ would be that successes are achieved whenever some minimal number of players cooperate ($\geq2$ say). Notice though, that such a choice introduces a second coordination problem over who cooperates and who defects. Efficiency requires the minimal number of cooperators, with the rest free riding. Changing the signal function in this way would therefore introduce a distributional asymmetry \citep[see][for an illustration that these additional dimensions lead to non-trivial problems]{wilson2020information}. Given our focus on understanding the effects from adding a new source of strategic uncertainty, we therefore chose an environment that did not introduce additional trade-offs/confounds.}  

Ignoring the game's scale and normalization (held constant in our experiments with $\Delta\pi=\$9$ and $\pi_0=\$11$) the repeated games we examine are summarized by three primitives: (i) The relative cost of cooperating, $x$. (ii) The number of players, $N$. (iii) The continuation probability, $\delta$. Our experiments will fix $\delta=\nicefrac{3}{4}$ in all but one diagnostic treatment in the extensions section. This leaves us with two key experimental parameters: the relative cost $x$ (actual cost $X=x\Delta\pi)$ and the number of participants $N$. 

In building a model of strategic uncertainty for arbitrary $N$, we will use a symmetric belief over the others' choices. That is, we assume each player chooses a mixture $p\cdot\EStrat{Grim}\oplus(1-p)\cdot\EStrat{All-D}$ over the two strategies.\footnote{For the $N$-player dilemma we define the grim-trigger strategy with imperfect-signals as:
$$
\EStrat{Grim}(h_t) =
\begin{cases}
    C & \text{if }t=1 \text{ or } h_t=((C,S),(C,S),\ldots,(C,S)) \\
    D & \text{otherwise.}
\end{cases}$$
} Our family of social dilemmas require cooperation from all $N$ players for everyone to get a success signal. Because of this, the strategic uncertainty reduces to the probability that the other $N-1$ players \emph{jointly} coordinate on the collusive strategy, 
$$Q\left(N\right)=\Prob{N-1 \text{ others all choose }\EStrat{Grim}}.$$ 
In every other case a failure will be received by at least $N-1$ players, and the punishment path will be triggered.

In an identical manner to the two-player case, the critical belief $Q^\star(N)$ is given by the point of indifference between the amount given up with certainty from a single round of cooperation, $x\cdot\Delta\pi$, and the continuation gain from collusion, $\tfrac{\delta}{1-\delta}\cdot\Delta\pi$, obtained with probability $Q(N)$. Indifference is therefore given by:
\begin{align*}
    Q^\star\left(N\right) =\frac{(1-\delta)}{\delta} x,
\end{align*}
where the RHS is identical to the two-player construction in (\ref{TwoPlayerBasin}) for $x=g=s$.

To complete the extension we therefore need to relate the joint-cooperation of the other $N-1$ players to the probability $p$ that each other player individually attempts to collude. However, even though we have specified the marginal belief distribution and assumed symmetry, we must still resolve the relationship between the joint and marginal distributions: the extent to which beliefs are correlated. In particular our design and extension will focus on two extreme models as we extend the model to $N$-players. The `standard' extension, where beliefs are fully independent; and an alternative/null-effect model where beliefs are perfectly correlated.\footnote{See \citet{casonCorBeliefs} for a clear example of correlated beliefs arising where independence would be the standard prediction.} Assuming perfect correlation for the other $N-1$ agents, joint and individual probabilities are identical, $Q\left(N\right)=p$, and so the extended critical belief (and correlated model outcome) is given by: 
\begin{equation}
\label{NplayerBasinCor}
p^\star_{\text{Corr.}}(x)=\dfrac{1-\delta}{\delta}\cdot x.
\end{equation}
In constrast, when the beliefs are independent we have $Q\left(N\right)={p}^{N-1}$, and the critical belief (and independent model extension) is:
\begin{equation}\label{NplayerBasinInd}
p^{\star}_{\text{Ind.}}(x,N)=\left(\dfrac{1-\delta}{\delta}\cdot x\right)^{\tfrac{1}{N-1}}\equiv \left(p^\star_{\text{Corr.}}(x)\right)^{\tfrac{1}{N-1}}.
\end{equation}
Obviously, when $N=2$ the basin measures in \eqref{NplayerBasinCor} and \eqref{NplayerBasinInd} are identical, dovetailing with the standard construction. However, for $N>2$ the two measures of strategic uncertainty are distinct, where the standard model extension under independence has strategic uncertainty vary both through the payoff parameter $x$ and the group-size $N$.

\section{Experimental Design\label{sec:design}}
The basin of attraction for always defect under independent beliefs serves as our measure of strategic uncertainty. Ceteris paribus, the greater the uncertainty on successful strategic coordination, the more likely the subject is to take refuge in the safer strategy---in the case of an RPD game, the stage-game Nash outcome of defecting. 

As outlined above, as we extend the environments to $N>2$, this opens up a question as to how strategic uncertainty is affected by $N$. If others' behavior is thought to be highly correlated, adding players but holding constant the payoffs will do little to affect the selected behavior in our games. If this is so, the null-effect $p^{\star}_{\text{Corr.}}$ measure of strategic-uncertainty will predict behavior. In contrast, in a more-standard extension where beliefs on the other players are independent, then $p^{\star}_{\text{Ind.}}$ will model the strategic uncertainty. Under this extension, we will be able to use shifts in $p^{\star}_{\text{Ind.}}$ to understand changes in the selected behavior. While one implication of this is that the addition of more players will ceteris paribus reduce collusion, a deeper implication of the model is to help us understand substitution effects across the two sources of strategic uncertainty, $x$ and $N$.

Our experimental design attempts to untangle the effects of strategic uncertainty. The aim of the design is to embed  comparative-static tests on the effect of $N$ (and so rule out the $p^{\star}_{\text{Corr.}}$ null-effect model), but also to examine the possible substitution effects by constructing perfect-substitutes treatments with the $p^{\star}_{\text{Ind.}}$ model. We achieve this through a series of experimental implementations of the $N$-player 2-action--2-signal repeated game outlined above. In particular, the first part of our design will use this family of games to distinguish and separate between the two extremes of independence and perfect correlation, leveraging the theoretical relationships derived above. 

While we cannot directly manipulate strategic uncertainty---as the basin-size measures are indirect, theoretical relationships derived from the primitives---equations \eqref{NplayerBasinCor} and \eqref{NplayerBasinInd} allow us to implicitly manipulate each measure through shifts in $x$ and/or $N$. Increases in $x$ increase the strategic uncertainty in both models: as higher costs of cooperation require a greater belief that the other(s) are cooperating. In contrast, increases to the number of players $N$ only increases strategic uncertainty for the independent basin-size measure, interacting with $x$ in a non-linear manner. 

Using equations \eqref{NplayerBasinCor} and \eqref{NplayerBasinInd},  the two notions can be varied in isolation from one another.  As such, it is possible to construct a $2\times 2$ design that orthogonally varies each strategic-uncertainty measure. Next, we outline our design, summarized in Table \ref{tab:TreatDesign}. 

Panel (A) of Table \ref{tab:TreatDesign} illustrates our first treatment dimension, which manipulates the payoff cost of cooperating $X=x\cdot\Delta\pi$, where $\Delta\pi=\$9$. The two values of $X$---a high temptation of $\$9$ (illustrated on the left, $x=1$), and a low temptation of $\$1$ (on the right, $x=\nicefrac{1}{9}$)---lead to two payoff environments over own-actions and the signals. The given stage-game payoffs capture what a participant would see on their screens.\footnote{See Figure \ref{fig:Screenshots} in the Online Appendix for representative screenshots.} 

\begin{table}[tb!]
    \caption{Experimental Design}
    \label{tab:TreatDesign}
    \centering
    \begin{tabular}{cccccc}
\toprule
\multirow[c]{2}{*}{\bf Panel A. \minitab[c]{ Stage-game \\ payoffs}}   & \multicolumn{2}{c}{$X=\$9$} & &\multicolumn{2}{c}{$X=\$1$ }\\ \cmidrule{2-3}\cmidrule{5-6}
 & $\sigma(a_{-i})=S$ & $\sigma(a_{-i})=F$ & & $\sigma(a_{-i})=S$ & $\sigma(a_{-i})=F$ \\ \midrule
  Coop., $\pi_i(C,\sigma)$ & $\$20$ & $\$2$ &  &  $\$20$ & $\$10$ \\
  Defect, $\pi_i(D,\sigma)$ & $\$29$ & $\$11$ & &  $\$21$ & $\$11$ \\
\midrule
\multirow[c]{2}{*}{\bf Panel B. \minitab[c]{All-D Basin \\ Size}} & \multicolumn{2}{c}{$X=\$9$ ($x=1$)} & &\multicolumn{2}{c}{$X=\$1$ ($x=\nicefrac{1}{9}$)}\\ \cmidrule{2-3}\cmidrule{5-6}
 & $N=2$ & $N=4$ & & $N=4$ & $N=10$ \\ \midrule
Cor. basin, $p^{\star}_{\text{Cor.}}(x)$ & $p^\star_0$ & $p^\star_0$&  & $p^\star_0-\Delta p^\star_{\text{Cor.}}$ & $p^\star_0-\Delta p^\star_{\text{Cor.}}$\\

& [0.33] & [0.33] & & [0.04] & [0.04] \\
  
Ind. basin, $p^{\star}_{\text{Ind.}}(x,N)$ 
& $p^\star_0$ & $p^\star_0+\Delta p^\star_{\text{Ind.}}$&  & $p^\star_0$ & $p^\star_0+\Delta p^\star_{\text{Ind.}}$\\
& [0.33] & [0.69] & & [0.33] & [0.69] \\
\cmidrule{1-6}
Sessions        & 3  & 3 & & 3 & 2 \\
Subjects        & 60 & 60 & & 72 & 60 \\
\midrule
\multirow[c]{2}{*}{\bf Panel C. \minitab[c]{ Meta-study \\ predictions}} & $p^\star_0$  &  \multicolumn{4}{c}{Marginal effect from:} \\ \cmidrule{3-6}
& [0.33] & \multicolumn{3}{c}{ Basin increase to [0.69]}&  Basin decrease to [0.04] \\
%&  & \multicolumn{3}{c}{Reduction of $\Delta p^\star_{\text{Cor.}}$} & Increase of $\Delta p^\star_{\text{Ind.}}$ \\ \midrule
%& %[0.333 ]  
%$[0.33]$ & \multicolumn{3}{c}{$[0.69]$} & [0.04] \\ \midrule
%& \multicolumn{3}{c}{($+\DeltaP{Ind.}=+0.36$)} & ($-\DeltaP{Corr.} =-0.29$) 
\midrule
Initial coop. ($t=1$) & $0.50$  & \multicolumn{3}{c}{$-0.26$}  & $+0.35$   \\
%Initial coop. ($t=1$)& $\mEst{0.495}{0.037}$  & \multicolumn{3}{c}{$\mEst{0.345}{0.070}$}  & $\mEst{-0.258}{0.036}$   \\
% & \SD{0.037} & \multicolumn{3}{c}{\SD{0.070}}& \SD{0.036}  \\
Ongoing coop. ($t>1$) & $0.37$ & \multicolumn{3}{c}{$-0.21$}  & $+0.50$  \\
% Ongoing coop. ($t>1$) & $\mEst{0.373}{0.035}$ & \multicolumn{3}{c}{$\mEst{0.495}{-0.347}$}  & $\mEst{-0.211}{0.026}$  \\ 
\bottomrule
\end{tabular}

\begin{tablenotes}
Meta-study prediction indicates the cooperation-rate estimation from the 2-player basin recovered from the treatment-clustered probits illustrated in Figure \ref{fig:MetaStudy}.
\end{tablenotes}
\end{table}

Our design also manipulates the number of players $N$, captured in the column headings of Panel (B) in Table \ref{tab:TreatDesign}. In total, we create four treatment environments, each defined by an $(N,X)$-pair, where we will refer to treatments with the directly varied parameter pair. In Panel (B), the rightmost four columns are headed by the chosen game variable, where the two rows of the table indicate how these choices affect the two basin-size measures of strategic uncertainty.  

To manipulate each basin-size measure separately, our design takes \Treat{N=2}{X=\$9} as its starting point. The values for both the independent and correlated basin-size measures are the same: $p^\star_0=0.33$. Holding the relative cooperation cost fixed and increasing the number of players to $N=4$ does not affect the correlated measure in \eqref{NplayerBasinCor}. However, a shift to $N=4$ increases the independent basin measure to $p^\star_0+\DeltaP{Ind.}=0.69$.

Now consider the manipulation of $X$. Comparing \Treat{N=4}{X=\$1} to \Treat{N=2}{X=\$9}, we hold constant the independent basin-size measure at $p^\star_0=0.33$. The shifts in both $X$ and $N$ have perfectly substituting effects in equation \eqref{NplayerBasinInd}. However, the same change in both variables under the correlated-basin measure has a substantial effect, where the change in $N$ does not provide an offsetting effect. As such, the correlated-basin measure is lowered to $p^\star_0-\DeltaP{Corr.}=0.04$. Finally, in the \Treat{N=10}{X=\$1} treatment we complete the $2\times2$ design over the two basin-size measures. Comparing \Treat{N=4}{X=\$1} to \Treat{N=10}{X=\$1} holds constant the correlated basin, which does not depend on $N$, at $p^\star_0-\DeltaP{Corr.}=0.04$. But the increased number of players does increase strategic uncertainty in the independent basin. In particular, our parameterization matches the independent basin sizes for \Treat{N=10}{X=\$1} and \Treat{N=4}{X=\$9} at $p^\star_0+\DeltaP{Ind.}=0.69$.

Through variation in the primitives $x$  and $N$, our design thereby generates four correlated/independent basin measure pairs with a $2\times2$ structure:\footnote{We note that our choices of $\Delta \pi=\$9$ and $\delta=\nicefrac{3}{4}$ were necessary choices here, as we aimed for exact-integer solutions for $N$ and $X$ for simplicity of the presentation. Our design over the basin measures is more-exactly given by:
$$ \left(p^{\star}_{\text{Corr.}},p^{\star}_{\text{Ind.}}\right) \in \NiceSetBig{3^{-1},3^{-3}} \times \NiceSetBig{3^{-1},3^{-1/3}}.$$
}
$$ \left(
p^{\star}_{\text{Corr.}},p^{\star}_{\text{Ind.}}\right) \in \NiceSetBig{p^\star_0,p^\star_0- \DeltaP{Corr.} } \times \NiceSetBig{p^{\star}_0, p^{\star}_0+ \DeltaP{Ind.} } :=\NiceSetBig{0.33,0.04} \times \NiceSetBig{0.33,0.69}.$$

The design therefore achieves the goal of orthogonal variation over the two basin measures. However, the parameterization was also chosen so that the shifts in each dimension would be expected to have quantitatively large effects. Through the \cite{dal2018determinants} meta-study we can generate level predictions for the behavioral effects of each change. Our design generates three basin-size measures: $p^\star_0=0.33$, $p^\star_0-\DeltaP{Corr.}=0.04$, and $p^\star_0+\DeltaP{Ind.}=0.69$. Using the probit models illustrated in Figure \ref{fig:MetaStudy} we leverage the meta-study to make \emph{quantitative} predictions for the cooperation rates under each basin-size measure. These predictions are indicated in Panel (C) of Table \ref{tab:TreatDesign}. 

Shifts in the basin-of-attraction under either model generate large changes in the predicted outcomes. The first column in Panel (C) indicates the baseline initial (ongoing) cooperation rate expected at $p^\star=0.33$ using the meta-data estimated relationship is 50 percent (37 percent).\footnote{All predictions based on late-session meta-study data (supergames 16--20), the same point in the sessions used to assess our hypotheses.} The next column pair then indicates the expected treatment effect for a shift in the strategic uncertainty level from $p^\star=0.33$ to either of the two alternative values. Panel (C) indicates that if either of the two strategic uncertainty measures captures and generalize behavior, then we would expect a large treatment effect on one of the two design dimensions.

We formalize the two competing hypotheses as:
\begin{corrhypothesis}
Cooperation increases as we decrease $X$, but there is no effect as we vary the number of players $N$. 
 \end{corrhypothesis}
\begin{indhypothesis}
Cooperation decreases as we increase $X$ and/or $N$. Moreover, the substitution effects between $X$ and $N$ indicate no effect on cooperation if we decrease $X$ and increase $N$ to hold constant the $p^\star_{\text{Ind.}}$ measure of strategic uncertainty.
\end{indhypothesis}

That is, consider the prediction if the standard independence-based extension of strategic uncertainty is the correct model. In treatments \Treat{N=2}{X=\$9} and \Treat{N=4}{X=\$1} the independent basin-size is 0.33, increasing to 0.69 in treatments \Treat{N=4}{X=\$9} and \Treat{N=10}{X=\$1}. If the strategic uncertainty relationship estimated from the two-player RPD meta-data is perfectly extrapolated then we should expect: (i) A reduction of 26 (21) percentage points in initial (ongoing) cooperation across the treatment pair, caused by the increase in strategic uncertainty. (ii)  A null effect on cooperation within each treatment pair, reflecting the designed perfect substitution across $X$ and $N$.\footnote{ Alternatively, under a null-effect from $N$, given by the correlated basin measure, the basin-size is reduced from 0.33 to 0.04 as we move between the \Treat{N=2}{X=\$9} and \Treat{N=4}{X=\$9} treatment pair and the \Treat{N=4}{X=\$1} and \Treat{N=10}{X=\$1} pair. The RPD prediction from the meta-study is then for an increase in the initial (ongoing) cooperation rate of 35 (50) percentage points (and again, a null effect within each pair).}

Notice that our hypotheses are silent with respect to which of the two outcome measures, initial and/or ongoing cooperation, we are supposed to match. The two measures have different interpretations---where initial cooperation captures intentions, and ongoing cooperation captures successful coordination. In the case of the two-player RPD, Figure \ref{fig:MetaStudy} shows that the basin size tracks both cooperation measures relatively well, and that the effects are hard to disentangle. Through $N$, we are able to generate additional variation in the theoretical relationship between initial and ongoing cooperation variables that separates the two relationships on another dimension. An advantage of our design is that it will allow us to better identify if either of the two measures is better predicted by either basin-size measure.

\subsection*{Experimental Specifics\label{sec:designspecifics}}
We use a between-subject design over the four distinct environments outlined in Table \ref{tab:TreatDesign}. Participants for each treatment were recruited from the undergraduate population at the University of Pittsburgh, participating in only one session. We recruited a total of 520 participants, with 252 for the first four treatments and 268 for the extensions that we outline in Section \ref{sec:extensions}. Three sessions were conducted for each environment, with an aim to recruit at least 20 participants per session---the one exception to this was the \Treat{N=10}{X=\$1} treatment, where we ran two sessions of 30.\footnote{In more detail, our design called for sessions to have at least 20 participants, but allowed us to recruit an additional group of size $N$ depending on realized show ups. For \Treat{N=10}{X=\$1} we instead opted to recruit 30 for each session so that we had at least three groups in each supergame.} Sessions lasted between 55 and 90 minutes with participants receiving an average payment of \$18.61.

Each session comprised 20 supergames, where supergames used a random termination chance of $1-\delta=\nicefrac{1}{4}$ after each completed round. However, supergame-lengths are matched by session across treatment. Participants were randomly and anonymously matched together across the 20 supergames in a stranger design.\footnote{All subjects received written and verbal instructions on the task and payoffs, where instructions are provided for readers in the Online Appendix.} The 20 supergames were divided into two parts of ten supergames.\footnote{Subjects received full instructions for the first part and were told they would be given instructions on part two after completing supergame ten. For the four between-subject treatments outlined in section \ref{sec:design}, part two was then identical to part one. Later in the paper we will outline a further set of treatments with a within-subject change across the parts. The design choice for two identical parts here allows for direct comparisons in first-half play.}  For final payment, one supergame from each part was randomly selected, where only the actions/signals from the last round in the selected supergame counted for payment.\footnote{This method is developed in \citet{sherstyuk2013payment} to induce risk neutrality over supergame lengths. Another benefit from this design choice is that there are no wealth effects within a supergame, and where history only matters as an instrument for others' future play.}

\section{Results\label{sec:results}}
In the first section we begin by describing the aggregate treatment-level cooperation rates. We postpone inferential tests of our two basin-extension hypotheses to the second section. The main finding here is that while neither extension contains all the relevant information for predicting initial cooperation, we do find more definitive results for ongoing cooperation. Overall, we conclude that the basin-size measure based on a standard independence assumption contains all relevant information for predicting successful coordination within the experimental supergames. 

\begin{table}[tb!]
    \centering
    \caption{Cooperation rates and basin-effect decomposition}
    \begin{tabular}{cccccc}
\toprule
\multirow[l]{2}{*}{\bf Panel A.\minitab[c]{\textbf{Action and} \\ \textbf{signal rates}}} & \multicolumn{2}{c}{$X=\$9$} & &\multicolumn{2}{c}{$X=\$1$}\\
\cmidrule{2-3}\cmidrule{5-6}
 & $N=2$ & $N=4$ & & $N=4$ & $N=10$ \\ 
 \midrule
  Initial coop. & $\mEst{0.503}{0.058}$ & $\mEst{0.035}{0.017}$ & & $\mEst{0.792}{0.042}$ & $\mEst{0.357}{0.055}$ \\ 
 Ongoing coop. & $\mEst{0.450}{0.055}$ & $\mEst{0.006}{0.003}$ & & $\mEst{0.409}{0.050}$ & $\mEst{0.184}{0.048}$  \\ 
 \cmidrule{2-6}
 Initial success & 0.503 & 0.000 & & 0.578 & 0.000 \\ 
 Ongoing success  & 0.450 & 0.000 & & 0.293 & 0.000  \\  
\midrule
\multirow[l]{2}{*}{\bf Panel B.\minitab[c]{\textbf{Cooperation} \\ \textbf{decomposition}}} & $p^\star_0$ &  \multicolumn{4}{c}{Marginal effect from: } \\ \cmidrule{3-6}
 &  & \multicolumn{2}{r}{Ind. basin increase to} &  \multicolumn{2}{r}{Corr. basin decrease to} \\
 & [0.33] &  \multicolumn{2}{r}{$p^\star_0+\DeltaP{Ind.}=[0.69]$ } & \multicolumn{2}{r}{$p^\star_0-\DeltaP{Corr.}=[0.04]$ }\\
 \midrule
 Initial coop. & $\mEst{0.464}{0.058}$ & \multicolumn{2}{c}{$\mEst{-0.395}{0.048}$} & \multicolumn{2}{c}{$\mEst{+0.357}{0.053}$ }  \\ 
 Ongoing coop. & $\mEst{0.366}{0.051}$ & \multicolumn{2}{c}{$\mEst{-0.293}{0.051}$} & \multicolumn{2}{c}{$\mEst{+0.115}{0.061}$ } \\ 
 \bottomrule
\end{tabular}

    \label{tab:Aggregate}
    
    \begin{tablenotes}
        Results are calculated using data from the last-five supergames. Cooperation rates present raw proportions (with subject-clustered standard errors). The cooperation decomposition runs two subject-clustered probits on the cooperation decision (initial and ongoing) where variables are dummies for a low correlated basin treatment ($X=\$1$, both $N$ values) and a high-independent--basin treatment ($X=\$9/N=4$ and $X=\$1/N=10$). Coefficients shown are the predicted level at just the constant (the $p^\star_0$ column) and the predicted cooperation changes from each estimated dummy.
    \end{tablenotes}
\end{table}

\subsection{Main Treatment Differences}
The top two lines of Panel (A) in Table \ref{tab:Aggregate} report cooperation rates broken out across the four treatments, where we separately report initial (the first round) and ongoing cooperation (all subsequent rounds). The averages in the table are for the last five supergames---late-session behavior, after subjects have amassed experience in the environment---though including all rounds generates similar results (see Table \ref{tab:CoopAllSG} in the Online Appendix). The results point to large cooperation shifts across both the cost to cooperating $X$ and the group size $N$. 

The initial cooperation rate in our \Treat{N=2}{X=\$9} treatment is 50.3 percent, essentially identical to the 50 percent cooperation rate predicted by the PD meta-study. However, holding constant the cooperation cost at $X=\$9$ and doubling the group size to four virtually eliminates cooperative behavior, with just 3.5 percent initial cooperation in \Treat{N=4}{X=\$9}. In low-temptation settings ($X=\$1$), groups of $N=4$ show highly cooperative behavior (79.2 percent for initial decisions), while groups of $N=10$ generate moderate cooperation rates (35.7 percent). The second row of Panel (A) indicates the ongoing ($t>1$) cooperation rate. Here the data indicates a decline in cooperation over the initial behavior in all treatments, though the quantitative effects are largest in the $\$X=1$ treatments.\footnote{In the Online Appendix, Table \ref{tab:history} further breaks out ongoing cooperation by the observed history in the previous round. The results indicate clear evidence that individual cooperation is highly conditional on successful coordination. However, strategies are significantly more-forgiving after failed cooperation at $X=\$1$ than $X=\$9$.} 
%[NOT SURE WHAT THIS WAS TRYING TO SAY] Overall, the magnitude of the decline for ongoing cooperation over initial is smaller in the $X=\$9$ treatments than the $X=\$1$ treatments.

The third and fourth rows of Table \ref{tab:Aggregate} present the fraction of participant rounds where a success signal was observed.\footnote{A success requires that the other $N-1$ participants jointly cooperate. Success is a direct function of group-level cooperation, where the \emph{expected} success rate with an independent cooperation rate $q$ is $q^{N-1}$. In two-player games, the success rate is identical to the cooperation rate. For the initial round the expected success rates (in the Table \ref{tab:Aggregate} column order) are: 0.503, $4.2\times 10^{-5}$, 0.497 and $9.5\times 10^{-5}$.} Focusing on success signals, similar patters emerge to ongoing cooperation, though with starker quantitative effects. While a success is the modal signal in the \Treat{N=2}{X=\$9} and  \Treat{N=4}{X=\$1} treatments, in the \Treat{N=4}{X=\$9} and  \Treat{N=10}{X=\$1} treatments we observe \textit{no} successes at all.\footnote{As success is a direct aggregate of individual level cooperation we do not report standard errors (where we also cannot calculate standard errors when there is no variation). However, the starkness of the effect with no successes when the independent basin-size is high make clear the underlying economic effects.}

Using just the raw averages in the top panel of Table 2, the evidence clearly falsifies the correlated-basin/null hypothesis on the effect of $N$, for both initial and ongoing cooperation. The experimental results indicate large shifts in behavior as we move $N$ as a comparative-static, fixing the value of $X$. Fixing $N=4$ we do find the comparative-static effect predicted by the correlated-basin measure as we move $X$, though this directional effect is also predicted by the independent basin.  

The evidence suggests that the independent-basin hypothesis fares slightly better. Both initial and ongoing cooperation clearly respond in the predicted direction as we shift either $X$ or $N$ in isolation. However, for initial cooperation, we do not find perfect substitution as we move both $x$ and $N$, as predicted by the independent-basin hypothesis. The hypothesis predicts no change in cooperation rates comparing either \Treat{N=2}{X=\$9} to \Treat{N=4}{X=\$1} or \Treat{N=4}{X=\$9} to \Treat{N=10}{X=\$1}. When this hypothesis is scrutinized using initial cooperation rates, we find large differences in either comparison.\footnote{The differences are economically large: 29 percentage points in the first comparison and 35 percentage points in the second.}  To see this finding from a different perspective, let us pool treatments at each value of $X$. That is, we compare the average initial cooperation rate pooling \Treat{N=2}{X=\$9} and \Treat{N=4}{X=\$9} to the average pooling \Treat{N=4}{X=\$1} and \Treat{N=10}{X=\$1}. Initial cooperation in the pooled treatments for $X=\$9$ and $X=\$1$ are 28.6 and 59.4 percent, respectively, and the difference is more than 30 percentage points. In other words, initial cooperation rates are inconsistent with the independent-basin hypothesis.

Results for ongoing cooperation though are substantially better for the independent basin, where \Treat{N=2}{X=\$9} and \Treat{N=4}{X=\$1} are not substantially different. We do note a difference between \Treat{N=4}{X=\$9} and  \Treat{N=10}{X=\$1}, %While we do find does persist over ongoing cooperation in the treatment pair where the independent basin measure indicates always defect is risk dominant, 
this is primarily driven by the very stark finding of near-zero cooperation in \Treat{N=4}{X=\$9}. We explore this further below. In the aggregate though, pooling the ongoing cooperation rates across $X$, we find similar rates at 22.8 percent for $X=\$9$ and 29.8 percent for $X=\$1$.

%leads to a significant difference in the initial cooperation rates, though with the sign of the difference different in each comparison. Even taking averages across the two comparisons by pooling the treatments at each value of $X$ yields a significant effect over initial cooperation, where the independent measure predicts a null-effect if the data is pooled this way. Initial cooperation in the pooled treatments for $X=\$9$ and $X=\$1$ are 28.6 and 59.4 percent, respectively. Results for ongoing cooperation though are substantially better for the independent basin, where \Treat{N=2}{X=\$9} are not \Treat{N=4}{X=\$1} substantially different. While a difference does persist over ongoing cooperation in the treatment pair where the independent basin measure indicates always defect is risk dominant, this is driven by near-zero cooperation in \Treat{N=4}{X=\$9}. Moreover, pooling the ongoing cooperation rates across $X$, and so averaging the differences, we find similar rates at 22.8 percent for $X=\$9$ and 29.8 percent for $X=\$1$.
%I rewrote the above a bit, as the comparison seemed convoluted.

\subsection{Evaluation of Independent- and Correlated-Basin Hypotheses}
Panel (B) in Table \ref{tab:Aggregate} provides a direct statistical evaluation of our two competing hypotheses. The table reports results of a probit model that assesses subjects' cooperation decisions using dummy variables for the 2$\times$2 design in Table \ref{tab:TreatDesign} Panel (B). The dummy covariates are an indicator for the $\Delta p^\star_{\text{Cor.}}$ decrease in the correlated basin-size (as we decrease $X$), and an indicator for the $\DeltaP{Ind.}$ increase in the independent basin-size (as we increase $N$ within each $X$).

Each row in Panel (B) provides the results from a separate estimation, one over initial cooperation, one over ongoing. The first column reports the estimated cooperation rate when both dummy variables are zero: essentially the cooperation rate for a game with a basin size of $p^\star_0=0.33$. The final column-pair then report the estimated marginal effect on the cooperation rate for each basin shock, holding the other constant. If either of the two basin hypotheses fully explains behavior, we would expect a significant estimate for the corresponding dummy and an insignificant effect on the other.

\begin{figure}[tb]
    \centering
    \subfloat[Initial cooperation]{
    \includegraphics[width=0.49\textwidth]{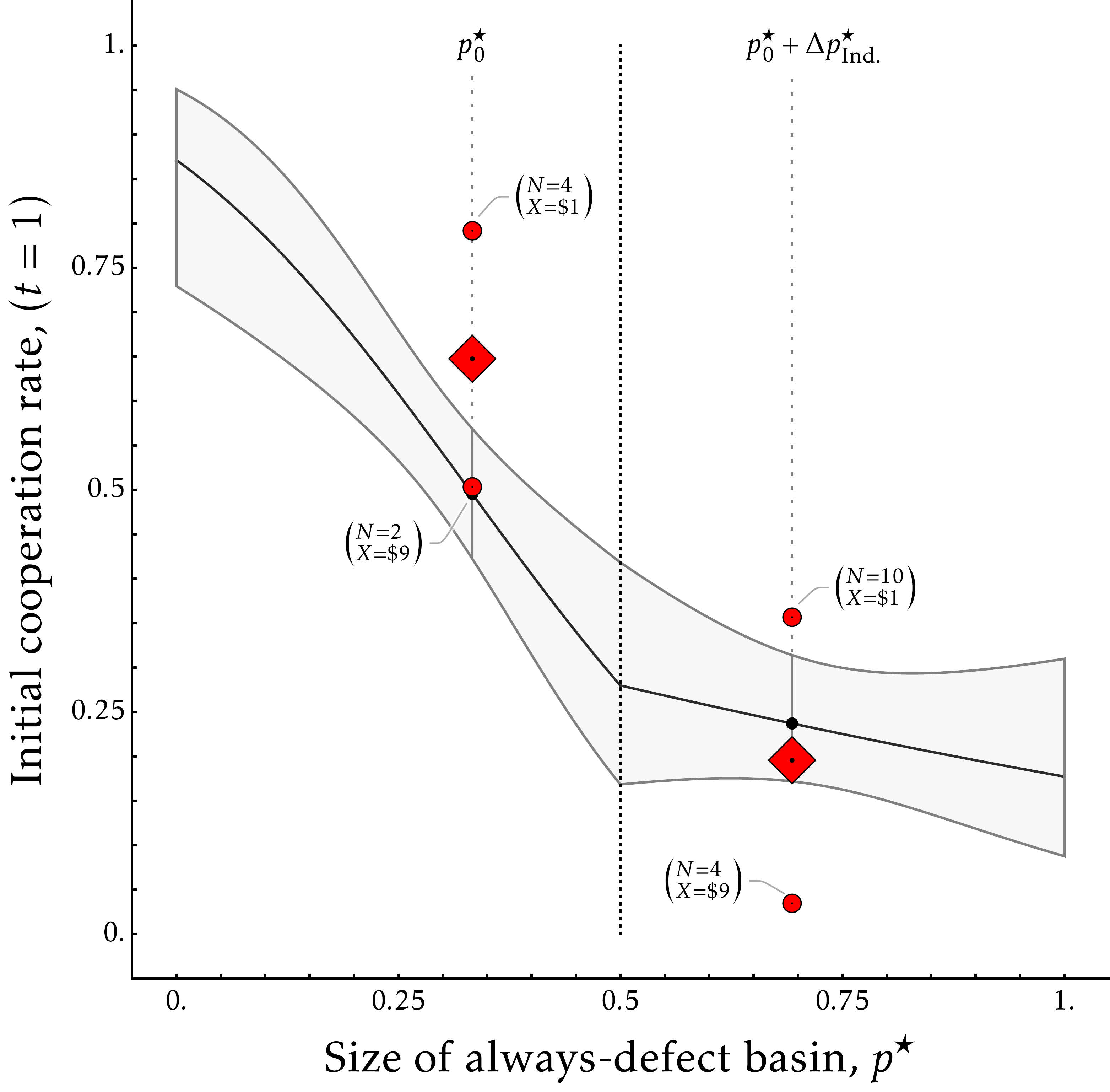}}
    \subfloat[Ongoing cooperation]{
    \includegraphics[width=0.49\textwidth]{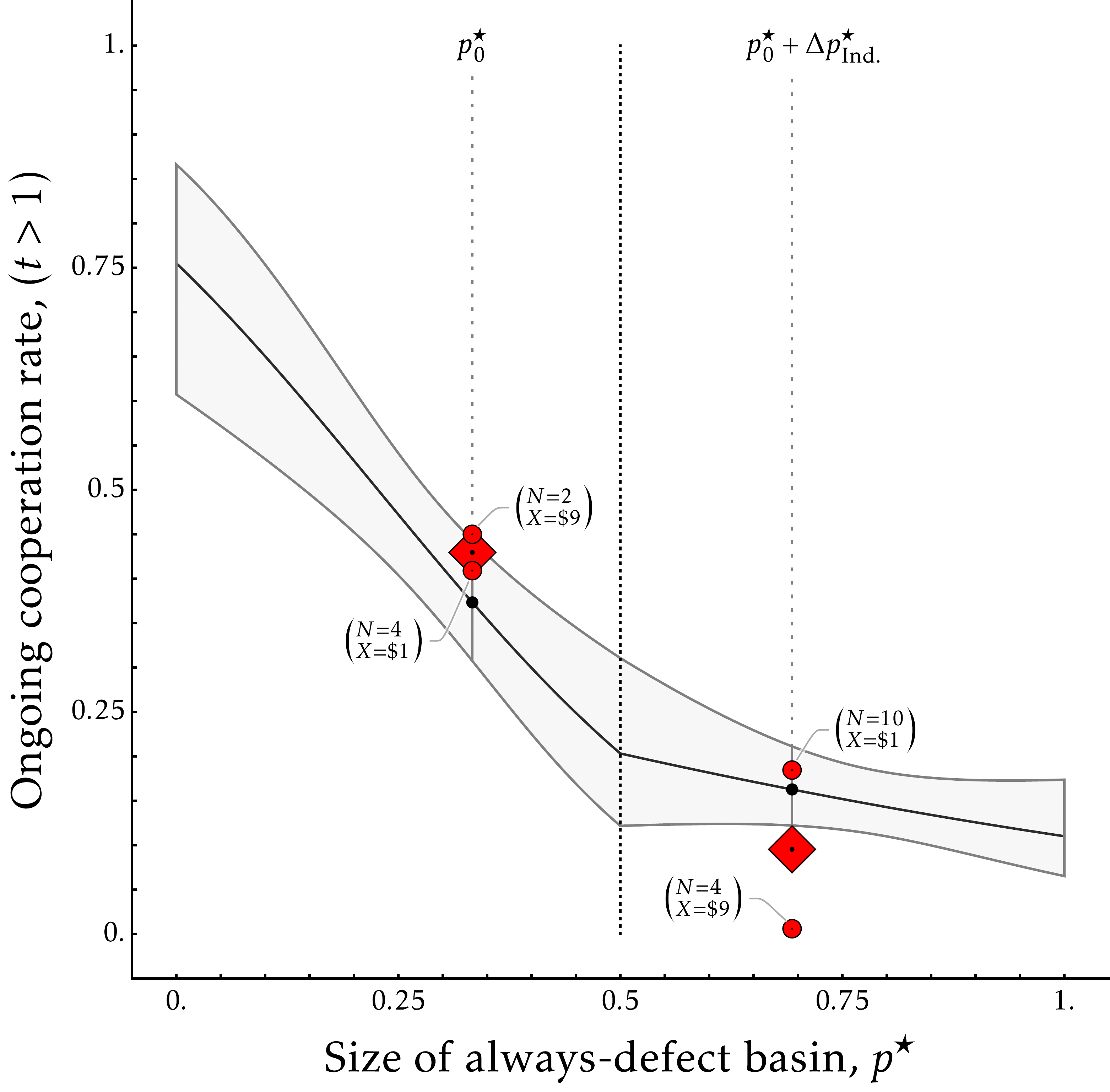}}
    \caption{Cooperation and the Independent Basin-Size Model}
    \label{fig:ResultsInd}
    
\begin{tablenotes}
Figures show pooled data by the independent basin (diamonds) while the separate treatments are illustrated as the surrounding circles. See Figure \ref{fig:ResultsCorr} in the appendix for an analogous figure for the correlated basin.
\end{tablenotes}
\end{figure}

The estimation procedure here is designed to directly parallel the probit-model used to generate predictions from the meta-study. The estimated cooperation rates at $p^\star_0=0.33$ is in fact quantitatively very close to the meta-study prediction in Panel (C) of Table \ref{tab:TreatDesign}. Where the meta-study predicts an initial (ongoing) cooperation of 49.5 (37.3) percent, our data at $p^\star_0=0.33$ indicates similar (and statistically inseparable) rates of 46.4 (36.6) percent. To illustrate this, in Figure \ref{fig:ResultsInd} we provide the fitted relationships from the meta study but where we overlay results from our four treatments (the smaller circles) using the independent basin-size on the horizontal axis. In addition we also indicate the results pooled over each value for the independent basin measure (the larger diamonds). While there is substantial divergence for initial cooperation in Panel (A), the quantitatively similar results for ongoing cooperation in Panel (B) are clear.\footnote{In Figure \ref{fig:ResultsCorr} in the Online Appendix we present an analogous figure organized under the correlated basin-size model, illustrating much poorer organization of the data, both relatively across the treatment comparisons, and quantitatively.}

Our tests of the two competing hypotheses are focused on the second and third columns of Panel (B) in Table \ref{tab:Aggregate}. If the independent (correlated) basin measure captures all relevant facets of behavior, we would expect a significantly negative (insignificant) estimate for the independent-basin increase (correlated-basin-decrease), and an insignificant (significantly negative) effect on the correlated (independent) basin-increase.

For initial cooperation, we find that changes to both basin-measures generate significant effects ($p<0.001$). The magnitudes of each estimated marginal effect are very similar, though each moving in opposite directions, as predicted. Since neither effect dominates, we conclude that both $X$ and $N$ contain information for predicting initial cooperation that is not fully absorbed by either basin measure.%\footnote{As our model is built under the assumption that payoffs reflect the preferences, one explanation for discrepancies in the initial cooperation behavior might be that for small values of $X$, other-regarding features of the preferences might dominate the raw effects.}% A: I ADDED THIS, BUT I'M NOT SURE ABOUT IT.

However, consistent with our descriptive presentation of the raw treatment rates in Table \ref{tab:Aggregate} Panel (A), the probit estimates for ongoing cooperation in Panel (B) do break towards the independent-basin construction. The coefficient on the increase in the independent basin is negative and significant ($p<0.001$), while the estimate on the decrease for the correlated basin is much smaller in magnitude and is not significant at the 5 percent level ($p=0.061$). Beyond just the qualitative directional effects, the quantitative change in ongoing cooperation under the independent-basin measure is close to the predicted effects from the relevant basin shifts that we would expect from the RPD meta-study data. That is, the predicted effect from the meta study for ongoing cooperation when the size of the basin increases from $p^\star_0=0.33$ to $0.69$ is a 21 percentage point drop (see Table \ref{tab:TreatDesign}, Panel C), where our estimates indicate a 29 percentage point drop.\footnote{In contrast, for a decrease to 0.04 we should expect an ongoing  cooperation-rate increase of 50 percentage points, where we instead we see 11.5 percent.}

As alluded to above, the differences on ongoing cooperation rates between our data and the out-of-sample prediction from the meta-study is driven by the stark (essentially boundary) behavior in the \Treat{N=4}{X=\$9} treatment. As illustrated in Figure \ref{fig:MetaStudy}(B) a two-player repeated game with a a basin of size $p^\star=0.69$ has a predicted ongoing cooperation rate of 16.3 percent, where we should be able to reject 11 percent cooperation at 95 percent confidence. While the \Treat{N=10}{X=\$1} treatment is close to the predicted rate (as are the two other treatments at $p^\star=0.33$), Figure \ref{fig:MetaStudy}(B) clearly indicates the \Treat{N=4}{X=\$9} treatment being significantly below the prediction. However, when the non-collusive strategy is risk dominant (when the independent basin size is greater than $\tfrac{1}{2}$), the argument made in the literature is that we should not expect substantial cooperation in this region. Where the model predicts low cooperation at this basin-size, the evidence from Treatment \Treat{N=4}{X=\$9} pushes towards this same conclusion, just in a starker way. Aside from the more-extreme coordination effects at \Treat{N=4}{X=\$9}, for the other three treatments the theoretically standard extension of the strategic uncertainty measure comes very close to \emph{quantitatively} predicting the ongoing cooperation level using the out-of-sample relationship estimated from the two-player RPD meta-data.

We summarize the main findings:
\begin{result}[Independent-Basin Measure]
    The independent-basin measure qualitatively organizes the results for ongoing cooperation, and in all but one treatment matches the quantitative level predictions. However, it does not contain all relevant information for predicting initial intentions to collude.
\end{result}

\begin{result}[Correlated-Basin Measure]
     Our data  is not consistent with the predictions from the correlated-basin hypothesis, neither for initial, nor ongoing cooperation. In particular, where the correlated basin predicts that behavior should ceteris paribus be unaffected by $N$ we instead find large cooperation decreases with increases to $N$.
\end{result}

\section{Extensions\label{sec:extensions}}
Our analysis so far has abstracted away other features of the coordination problem to focus on the pure effects from the primitives of the strategic game. In this section we consider two extensions---with relevance both inside and outside the laboratory---that allow us to study possible limitations of the strategic-uncertainty model to predict changes in equilibrium selection.

First, we consider the extent to which beliefs on others' collusive behavior may be distorted by prior experience. While a policy change can change market primitives and the strategic-uncertainty measure, the underlying variable in this model is beliefs on others' collusion. It seems plausible that beliefs may be driven by experience before any change in the primitives, and so the model may fare poorly at predicting changes within a population. For example, if a player has engaged extensively with the \emph{same} population of market participants under a status quo that led to non-collusive behavior, their beliefs on others acting in this way may be sticky, and therefore unresponsive to shifts in the primitives that have large effects in the strategic uncertainty model. Our treatments in the previous section used a between-subjects design. Identification relied on comparisons of late-session behavior across different populations, each with experience under a fixed environment. In a modified treatment pair we examine the effects of varying the number of players $N$ within the same population, so that participants have experience at two distinct values. In this extension we show that outcomes do not exhibit long-run stickiness, where the between-subject results detailed previously are not substantially dissimilar when we examine them within-subject.

In a second set of extensions, we examine the strategic uncertainty mechanism underlying the basin-size selection device. In particular we examine the extent to which our results are affected by the possibility of explicit coordination, holding constant $X$ and $N$. Here we seek to mirror an empirical finding that when collusion in industries is detected, it is often accompanied by evidence of explicit collusion---despite the illegality of such meetings.\footnote{See \citet{marshall2012economics} for a more comprehensive treatment.} In one treatment  we show that once explicit collusion is allowed for, neither the independent nor the correlated basin measures does a good job of predicting collusive behavior levels. Once parties can explicitly collude we find very high levels of sustained cooperation. This suggests that indeed uncertainty over the other strategic choices is a main driver of behavior in our main treatments. However, the extremeness of the effect once communication is allowed for raises a question over the extent to which explicit collusion might lead to high cooperation rates even when collusive outcomes are not equilibrium. To examine this, we show that there a clear limits on what explicit collusion can achieve is defined by the theoretical existence of collusive equilibria.

\subsection{Between vs. Within  Identification}
The motivating idea for our first extension is that in many settings of interest the policy-relevant comparative static is being varied within a population. However, if agents have strong beliefs about others due to prior experiences, it may be that our theoretical construction lacks bite at predicting outcomes as the policy shifts. If selected equilibria are very sticky within a population, then more-standard assumptions maintaining the equilibria across the counterfactual may have greater validity. For example, within the experiment if a participant's experience with others is that they play the stage-game defection action every period, then this belief can persist despite a policy shift that makes collusion easier. %This is particularly true when the policy variable is simply a change in the number of players $N$ in our environment. Given the same exact stage game (fixed $X$), and a matched participant who was previously defecting, 
%A change in the selected equilibrium across the change in policy requires participants to be beliefs to be bootstrapped upward with the policy variable, ignoring the past experiences. 

Ideally, we would introduce a primitive change within a supergame---for example, a move from $N=4$ to $N=2$, where the matched player after the modification is one of the matched participants from before. In exploring potential designs for this, we were not satisfied that they would produce clear results. First, it is well-documented that repeated-game environments require several supergames of experience for participants to internalize the environment \citep{bo2005cooperation}. While implementing a surprise change in $N$ as a mid-supergame manipulation would mirror an outside-the-laboratory consolidation, this would provide a single supergame observation, and require substantial explanation across the surprise. An alternative design choice could implement a change in $N$ with some probability within each supergame. However, any observed effects would then be confounded with the expectations over the primitive change (and greater complication in the instructions) and would no longer comparable to our between-subject treatments.

Given the potential confounds with other designs we instead opted---certainly as a first approach--- for a design with a surprise one-time shift in the number of players $N$, but where this occurred in a fixed session-level population. Holding constant the cooperation cost at $X=\$9$, we initially set a value of $N$ (either two or four) for the first ten supergames. We then change the value of $N$ for the last ten supergames (to either four or two, respectively).

This led to two further experimental treatments, one with \Treat{N=2}{X=\$9} in the first half, and \Treat{N=4}{X=\$9}  in the second; and the converse treatment from \Treat{N=4}{X=\$9} in the first, to \Treat{N=2}{X=\$9} in the second half. In both treatments, the change in $N$ comes as a surprise: subjects know there is a second part, but do not receive instructions on it until supergame ten concludes.\footnote{Our between-subject treatments also divided the session into two parts, except that once subjects reached the second half of the session they were told that part two was identical to part one.} In terms of the standard strategic-uncertainty model this creates a shift across the session from a low basin-size of 0.33 when $N=2$, and a high basin-size of 0.69 when $N=4$. In particular, this is a change in $N$ where our between-subject design indicates a substantial treatment effect. Given that we hold constant $X=\$9$, for simplicity we label the treatments as $2\rightarrow 4$ and $4\rightarrow 2$, for the first-half to second-half shifts.

\begin{figure}[tb!]
\centering
    \subfloat[Between vs. Within ($X=\$9$)]{
    \includegraphics[width=0.49\textwidth]{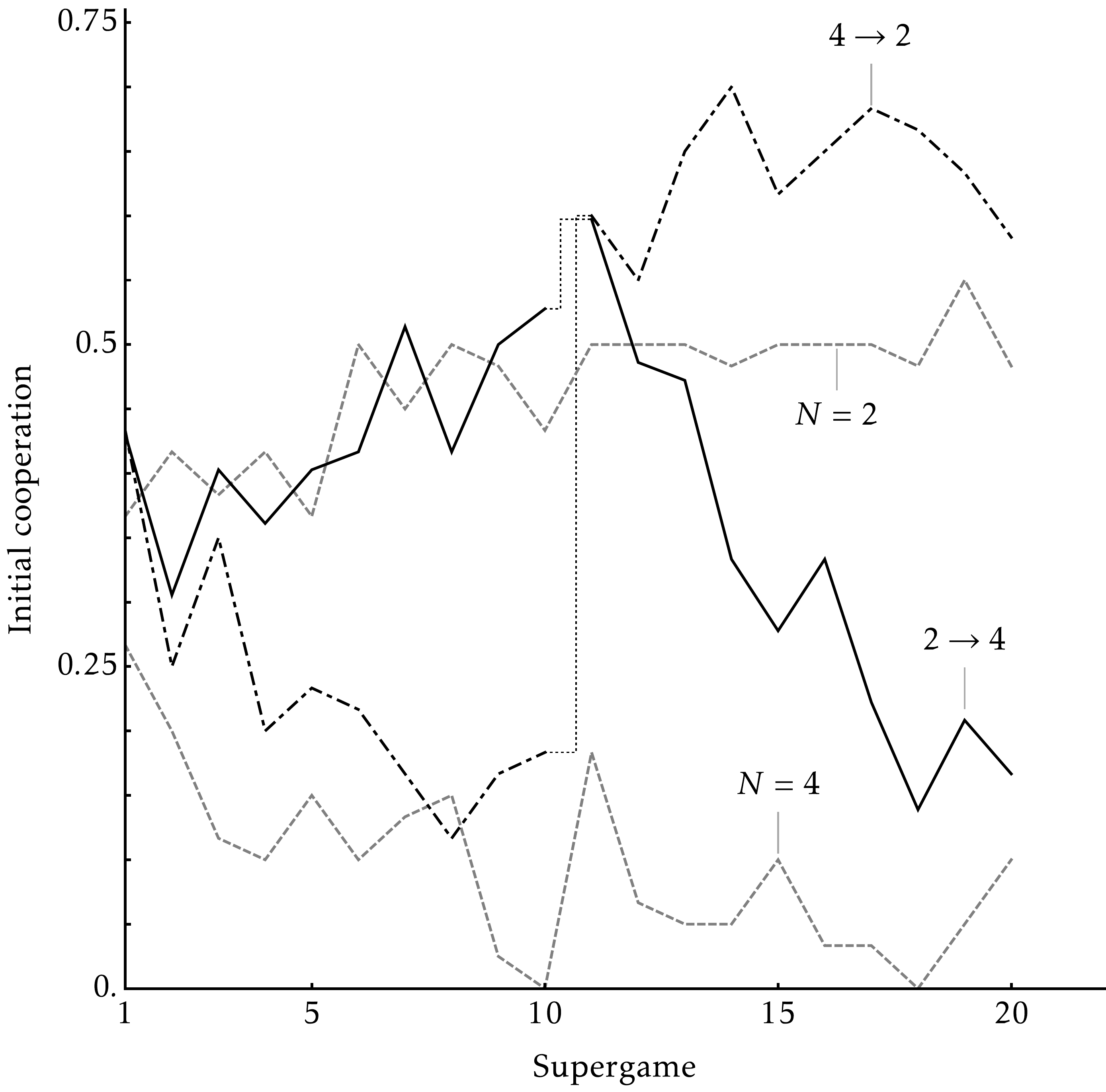}}
    \subfloat[Explicit vs. Implicit (\Treat{N=4}{X=\$9})]{
    \includegraphics[width=0.49\textwidth]{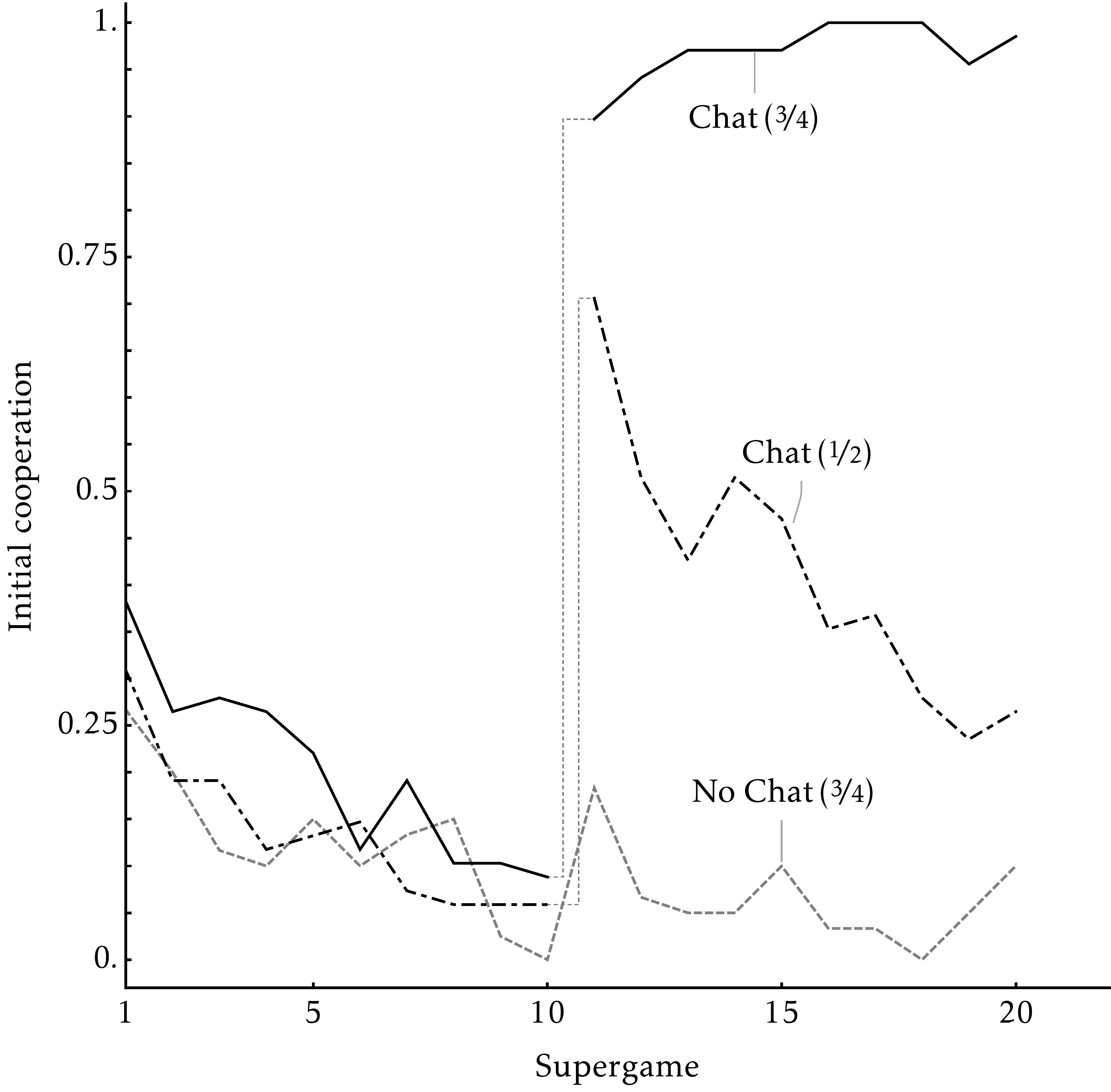}}
    \caption{Initial cooperation rates in extensions (by supergame)}
    \label{fig:Extensions}
\end{figure}

In Figure \ref{fig:Extensions}(A) we present the average initial cooperation-rates across the session's twenty supergames. The between-subject treatments with $N=2$ and $N=4$ are indicated by the two gray dashed lines (separately labeled), while the within-subject treatment results are represented by two black lines: a solid line for the $2\rightarrow 4$ treatment, and a dash-dotted-line for the $4\rightarrow 2$ treatment. The figure illustrates the substantial between-subject effect, with more cooperation in \Treat{X=\$9}{N=2} over \Treat{X=\$9}{N=4} for all twenty supergames. Our within and between sessions are identical for the first ten supergames, where the figure indicates replicated effects. Pooling the between and within treatments with $N=2$ in supergames 6--10 the initial cooperation rate is 47.4 percent. In contrast, the pooled cooperation rate for $N=4$ is just 13.9 percent.\footnote{Testing the initial cooperation rate differences in supergames 6--10 over $N$ (so across the between and within sessions with identical treatment at this point) we find $p=0.150$ for $N=2$ and $p=0.981$ for $N=4$ from $t$-tests for a level difference, and $p=0.353$ for a joint test.}

As we move into supergames 11--20, the number of players matched in each supergame changes for our within treatments. Figure \ref{fig:Extensions}(A) indicates the immediate shift in behavior as the primitive changes as the two vertical dotted lines. For the $2\rightarrow4$ treatment (the black solid line) initial cooperation levels remain fairly high after the shift from $N=2$ to $N=4$. In fact, cooperation in the first four-player interaction (supergame 11) actually exhibits an increase to 59.7 percent from the 53.0 percent from the last two-player interaction (supergame 10). However, while there is no immediate cooperation drop-off, and thus stickiness, as subjects gain experience at $N=4$ the cooperation rate falls rapidly, reaching 16.7 percent by supergame 20. In contrast, moving in the other direction from $N=4$ to $N=2$ (the black dash-dot line) we find an immediate jump with the primitive shift. Where initial-round cooperation in supergame 10 with four-players is 18.3 percent, the reduction to $N=2$ pushes this rate up to 60.0 percent for supergame 11. The immediate jump in behavior is then sustained across the remaining supergames, with 58.3 percent cooperation in supergame 20.

Inspecting the session time-series illustrated in Figure \ref{fig:Extensions}(A) it is clear that the there is little evidence for the hypothesis that equilibrium selection is sticky in the long-run under a within-population shift in $N$. Despite exposure to the alternative environment in the first half, longer-run behavior in the second-half is not substantially dissimilar from the between-subject levels. This is indicated by the close proximity of the two black/gray line pairs at supergame 20, and relative distance from the other pair.

In Online Appendix B we provide a more detailed like-with-like comparison of the between-subject and within-subject results. These more-detailed findings indicate no difference with the between-subject results as we move from $2\rightarrow 4$. However, in opposition to the hypothesis that the selected equilibrium is sticky, we actually find a significant \emph{increases} in the response to $N$ over the between analysis as we move from $4\rightarrow 2$. We summarize the result:

\begin{result}[Between vs. Within]
Switching the identification to within does not substantially change our qualitative results, finding no evidence that the selected equilibrium is sticky in the long run as we shift a primitive within the population. If anything, our within-subject identification shows a larger shift than the between-subject results as we decrease $N$.  
\end{result}

\subsection{Explicit Correlation}
The focus so far, and for the RPD literature more generally, is an examination of \emph{implicit} coordination. Our results thus far suggest that a model of strategic uncertainty extended via independence is relatively successful at organizing the data. This is particularly so for ongoing cooperation. In contrast, a measure based on perfectly correlated play fails to predict the substantial effect of $N$ for both initial and ongoing behavior.

These results are therefore a validation of a standard extension of the basin of attraction notion. However, correlation across participants' beliefs becomes much more plausible when participants have access to a channel through which to explicitly coordinate. In an extension to our previous experiments we now examine the extent to which multiple parties successfully collude when given a coordination device.

To this end, we designed a further treatment pair that modifies our environment with the starkest coordination on the non-collusive outcome, the \Treat{N=4}{X=\$9} treatment. In the first chat treatment, the first 10 supergames were identical to the \Treat{N=4}{X=\$9} treatment, but we introduce pre-supergame chat between the four matched players in supergames 11--20. A second chat treatment is identical to the first in terms of the timing for when we introduce the chat coordination device; the difference is that we reduce the continuation probability to $\delta^\prime=\nicefrac{1}{2}$ for the entire session. The designed effect of this change is that while the stage-game payoffs and number of players is held constant, at this value of $\delta$ the grim trigger is now only a knife-edge SPE, where the critical belief $p^\star(\delta^\prime)=1$ under both \eqref{NplayerBasinCor} and \eqref{NplayerBasinInd}. The $\delta=\tfrac{1}{2}$ treatment therefore serves as a test for whether explicit coordination can implement outcomes that are not supportable as a robust equilibrium (that is, with arbitrarily small trembles in behavior).

\begin{table}[tb!]
    \centering
    \caption{Cooperation: Implicit vs. Explicit}
    \label{tab:Chat}
    \begin{tabular}{ccccc}
\toprule  & Implicit & &\multicolumn{2}{c}{Explicit}\\ \cmidrule{2-2}
\cmidrule{4-5}  & $\text{NoChat}(\nicefrac{3}{4})$ &  & $\text{Chat}(\nicefrac{3}{4})$ & $\text{Chat}(\nicefrac{1}{2})$\\ 
\midrule 
%Corr Basin & 0.333 & 0.333 & 1.00\\ 
%Ind. Basin & 0.693 & 0.693 & 1.00\\ \cmidrule{2-4} 
Initial coop. & $\mEst{0.035}{0.017}$ & & $\mEst{0.988}{0.007}$ & $\mEst{0.300}{0.037}$ \\ 
Ongoing coop. & $\mEst{0.006}{0.003}$ & & $\mEst{0.806}{0.030}$ & $\mEst{0.044}{0.018}$\\ 
\cmidrule{2-5} Initial success & 0.000 & & 0.971 & 0.094\\ 
Ongoing success & 0.000 & & 0.756 & 0.002\\ 
\bottomrule 
\end{tabular}
    
    \begin{tablenotes}
        All treatments have $x=\$9, N=4$, where $\text{NoChat}(\nicefrac{3}{4})$ is from the core $2\times2$ between-subject design outline in section \ref{sec:results}.   Data taken from last five supergames, participant-clustered standard errors in parentheses.
    \end{tablenotes}
\end{table}

In an online supplement we provide chat logs for the interested reader; however,  we here focus on the of effects of chat on cooperation levels. In Figure \ref{fig:Extensions}(B) we  plot the average initial cooperation rates across sessions (with the \Treat{N=4}{x=\$9} treatment provided as a baseline, labeled here as NoChat($\nicefrac{3}{4}$)). Late-session cooperation and success rates (assessed in supergames 16--20 with subject-clustered standard errors) are provided in Table \ref{tab:Chat}.

Our first chat treatment delivers unequivocal results: providing pre-play communication at $\delta=\nicefrac{3}{4}$ takes the near-zero cooperation rate in the absence of chat to almost complete cooperation. While ongoing cooperation drops slightly from the very high initial-cooperation levels, the large majority of supergames exhibit coordination by all four participants on the efficient/collusive outcome. Such high levels of cooperation with communication are inconsistent with the predictions of either model of equilibrium selection for these primitives. This suggests that once explicit coordination devices are allowed for and strategic uncertainty is reduced, the independent model--that captured behavior when collusion is tacit--is no longer helpful. 

However, at the $\delta=\nicefrac{1}{2}$ boundary for dynamic strategies to be incentive compatible, even with pre-play communication participants find it hard to coordinate. While initial cooperation levels are substantially higher than the treatment without chat at 30.0 percent, ongoing cooperation falls to just 4.4 percent, where successful joint-cooperation across an entire group in an ongoing round becomes exceedingly rare at 0.2 percent. The findings therefore indicate that for explicit communication to play a role, clear incentives for collusion do need to be present.

Given the very stark findings here, we simply summarize the extension result:
\begin{result}[Implicit vs. Explicit]
    Explicit coordination leads to very high cooperation levels with multiple players, in a setting where implicit cooperation achieves near-zero cooperation. However, in the limiting case, where cooperation is a knife-edge SPE outcome, even pre-play chat fails to support cooperation in our experiments.
\end{result}

\section{Conclusion\label{sec:conclusion}}
Our paper examines equilibrium selection in repeated games, and the extent to which it can be predicted with a model of strategic uncertainty. We leverage a model of equilibrium selection that rationalizes behavior in data from repeated prisoner's dilemma (RPD) experiments and design an experiment to stress test this specific theoretical model. The predictive model works by mediating the effects from multiple-primitives into a single dimension, strategic uncertainty. As such, even for rich counterfactual policies with many changes to the setting, the model can still generate a directional prediction. The experimental design builds on this motivation for the model, introducing a novel source of strategic uncertainty that has not been studied in the RPD setting (the number of players) while also manipulating a payoff parameter. We can thus change both sources of strategic uncertainty simultaneously and study the extent to which the evidence is consistent with the predictions of the selection model. Our main finding is that the model of equilibrium selection can indeed be used as a device for understanding successful coordination on the collusive outcome. In particular, the model performs well in trading-off the competing effects from the two distinct sources of strategic uncertainty.

After illustrating the theoretical power of the model for implicit coordination, we turn our attention to two application-motivated extensions that probe its limitations. In the first, we are motivated by the extent to which prior history and experience could make the equilibria ``sticky,'' even where the model suggests a change. To do this, we study the extent to which our findings still hold when treatment-variable manipulations takes place within the same population, as opposed to between populations in our main study. The results indicate that eve when the participants experience a treatment-variable shift, the model still continues to predict the longer run outcomes. While we do find some hysteresis in the short-run response to a within-population policy change---initial stickiness in behavior in one direction and a large immediate response to the change in the other---behavior after some experience  under the new parameters is not distinct from the between-population treatments. As such, while a plausible limitation of the model was the effect of prior experience, these results suggest that the model prediction fares better than the more-standard assumption of maintaining a selected equilibrium across a policy change.

Finally, in a second extension, we examine the potential effects from explicit collusion. Shifting the experimental environment by providing an explicit coordination device (pre-play chat) we examine a necessary limitation on the model if the underlying strategic uncertainty mechanism is correct. Our main finding here is that once we allow for explicit coordination then the selection-model prediction is very far from observed behavior. In fact, where the model suggests very low levels of cooperation, actual behavior is at the other extreme once coordination devices are present. The evidence from the second extension therefore indicates both that strategic uncertainty plays a clear role, but also that the selection model based upon it is entirely inappropriate for predicting behavior when explicit collusion is suspected.

Though our results are encouraging, for the theoretical selection model we examine to be useful in applications further tests are needed. First, as we point out when describing our experimental design, there are many ways in which expanding the multiple players might impact the game.Because we do not want to introduce further coordination issues, we focus on a setting where collusion requires joint cooperation form all of the agents. But in other interesting environments cooperation among a subset of the agents may be enough for reaching an efficient outcome. Such environments introduce further distributional coordination problems, over who cooperates and who free-rides. Further research is needed to evaluate how and if models of equilibrium selection might be extended to such problems. Outside from the role of multiple players, there are also outstanding questions with regard to how asymmetries for the players---in terms of actions, payoffs, and the sequentiality of moves---should be modeled.

From another perspective, our results also outline that there is room to evaluate the role for equilibrium selection model in environments with explicit collusion. Our findings suggest one clear limit on where explicit collusion can play a role: the existence of a collusive equilibrium. However, beyond this point, not much is known about the conditions for which explicit communication between the parties is enough for collusion. Another direction for future research is to try to understand if there are ways to manipulate the primitives to understand when explicit collusion will and will not be successful.

\bibliographystyle{lit/te}
\bibliography{lit/lit.bib}

\begin{thebibliography}{46}
\newcommand{\enquote}[1]{``#1''}
\providecommand{\natexlab}[1]{#1}
\providecommand{\url}[1]{\texttt{#1}}
\providecommand{\urlprefix}{URL }
\providecommand{\bibAnnoteFile}[1]{%
  \IfFileExists{#1}{\begin{quotation}\noindent\textsc{Key:} #1\\
  \textsc{Annotation:}\ \input{#1}\end{quotation}}{}}
\providecommand{\bibAnnote}[2]{%
  \begin{quotation}\noindent\textsc{Key:} #1\\
  \textsc{Annotation:}\ #2\end{quotation}}

\bibitem[{Agranov et~al.(2016)Agranov, Frechette, Palfrey, and
  Vespa}]{agranov2016static}
Agranov, Marina, Guillaume Frechette, Thomas Palfrey, and Emanuel Vespa (2016),
  \enquote{Static and dynamic underinvestment: An experimental investigation.}
  \emph{Journal of Public Economics}, 143, 125--141.
\bibAnnoteFile{agranov2016static}

\bibitem[{Battaglini et~al.(2012)Battaglini, Nunnari, and
  Palfrey}]{battaglini2011legislative}
Battaglini, Marco, Salvatore Nunnari, and Thomas~R Palfrey (2012),
  \enquote{Legislative bargaining and the dynamics of public investment.}
  \emph{American Political Science Review}, 106, 407--429.
\bibAnnoteFile{battaglini2011legislative}

\bibitem[{Battaglini et~al.(2016)Battaglini, Nunnari, and
  Palfrey}]{battaglini2016dynamic}
Battaglini, Marco, Salvatore Nunnari, and Thomas~R Palfrey (2016), \enquote{The
  dynamic free rider problem: A laboratory study.} \emph{American Economic
  Journal: Microeconomics}, 8, 268--308.
\bibAnnoteFile{battaglini2016dynamic}

\bibitem[{Battalio et~al.(2001)Battalio, Samuelson, and
  Van~Huyck}]{battalio2001optimization}
Battalio, Raymond, Larry Samuelson, and John Van~Huyck (2001),
  \enquote{Optimization incentives and coordination failure in laboratory stag
  hunt games.} \emph{Econometrica}, 69, 749--764.
\bibAnnoteFile{battalio2001optimization}

\bibitem[{Berry et~al.(2017)Berry, Coffman, Hanley, Gihleb, and
  Wilson}]{berry2017assessing}
Berry, James, Lucas~C Coffman, Douglas Hanley, Rania Gihleb, and Alistair~J
  Wilson (2017), \enquote{Assessing the rate of replication in economics.}
  \emph{American Economic Review}, 107, 27--31.
\bibAnnoteFile{berry2017assessing}

\bibitem[{Blonski and Spagnolo(2015)}]{blonski2015prisoners}
Blonski, Matthias and Giancarlo Spagnolo (2015), \enquote{Prisoners’ other
  dilemma.} \emph{International Journal of Game Theory}, 44, 61--81.
\bibAnnoteFile{blonski2015prisoners}

\bibitem[{Brandts and Cooper(2006)}]{brandts2006change}
Brandts, Jordi and David~J Cooper (2006), \enquote{A change would do you
  good.... an experimental study on how to overcome coordination failure in
  organizations.} \emph{American Economic Review}, 96, 669--693.
\bibAnnoteFile{brandts2006change}

\bibitem[{Cason(2008)}]{cason2008price}
Cason, Timothy~N. (2008), \enquote{Price signaling and `cheap talk' in
  laboratory posted offer markets.} \emph{Handbook of Experimental Economics
  Results}, 1, 164--169.
\bibAnnoteFile{cason2008price}

\bibitem[{Cason et~al.(2020)Cason, Sharma, and Vadovi\v{c}}]{casonCorBeliefs}
Cason, Timothy~N., Tridib Sharma, and Radovan Vadovi\v{c} (2020),
  \enquote{Correlated beliefs: Predicting outcomes in $2\times2$ games.}
  \emph{Games \& Economic Behavior}, 122, 256--276.
\bibAnnoteFile{casonCorBeliefs}

\bibitem[{Compte et~al.(2002)Compte, Jenny, and Rey}]{compte2002capacity}
Compte, Olivier, Frederic Jenny, and Patrick Rey (2002), \enquote{Capacity
  constraints, mergers and collusion.} \emph{European Economic Review}, 46,
  1--29.
\bibAnnoteFile{compte2002capacity}

\bibitem[{Cooper and K{\"u}hn(2014)}]{cooper2014communication}
Cooper, David~J. and Kai-Uwe K{\"u}hn (2014), \enquote{Communication,
  renegotiation, and the scope for collusion.} \emph{American Economic Journal:
  Microeconomics}, 6, 247--78.
\bibAnnoteFile{cooper2014communication}

\bibitem[{Dal~B{\'o}(2005)}]{bo2005cooperation}
Dal~B{\'o}, Pedro (2005), \enquote{Cooperation under the shadow of the future:
  experimental evidence from infinitely repeated games.} \emph{American
  Economic Review}, 95, 1591--1604.
\bibAnnoteFile{bo2005cooperation}

\bibitem[{Dal~B{\'o} and Fr{\'e}chette(2011)}]{DalBoFrechette2011}
Dal~B{\'o}, Pedro and Guillaume~R Fr{\'e}chette (2011), \enquote{The evolution
  of cooperation in infinitely repeated games: Experimental evidence.}
  \emph{American Economic Review}, 101, 411--429.
\bibAnnoteFile{DalBoFrechette2011}

\bibitem[{Dal~B{\'o} and Fr{\'e}chette(2018)}]{dal2018determinants}
Dal~B{\'o}, Pedro and Guillaume~R Fr{\'e}chette (2018), \enquote{On the
  determinants of cooperation in infinitely repeated games: A survey.}
  \emph{Journal of Economic Literature}, 56, 60--114.
\bibAnnoteFile{dal2018determinants}

\bibitem[{Davis(2002)}]{davis2002strategic}
Davis, Douglas~D (2002), \enquote{Strategic interactions, market information
  and predicting the effects of mergers in differentiated product markets.}
  \emph{International Journal of Industrial Organization}, 20, 1277--1312.
\bibAnnoteFile{davis2002strategic}

\bibitem[{Deneckere and Davidson(1985)}]{deneckere1985incentives}
Deneckere, Raymond and Carl Davidson (1985), \enquote{Incentives to form
  coalitions with bertrand competition.} \emph{RAND Journal of Economics},
  473--486.
\bibAnnoteFile{deneckere1985incentives}

\bibitem[{Devetag and Ortmann(2007)}]{devetag2007and}
Devetag, Giovanna and Andreas Ortmann (2007), \enquote{When and why? a critical
  survey on coordination failure in the laboratory.} \emph{Experimental
  economics}, 10, 331--344.
\bibAnnoteFile{devetag2007and}

\bibitem[{Dufwenberg and Gneezy(2000)}]{dufwenberg2000price}
Dufwenberg, Martin and Uri Gneezy (2000), \enquote{Price competition and market
  concentration: an experimental study.} \emph{International Journal of
  Industrial Organization}, 18, 7--22.
\bibAnnoteFile{dufwenberg2000price}

\bibitem[{Farrell and Shapiro(1990)}]{farrell1990horizontal}
Farrell, Joseph and Carl Shapiro (1990), \enquote{Horizontal mergers: an
  equilibrium analysis.} \emph{American Economic Review}, 107--126.
\bibAnnoteFile{farrell1990horizontal}

\bibitem[{Fonseca and Normann(2012)}]{fonseca2012explicit}
Fonseca, Miguel~A and Hans-Theo Normann (2012), \enquote{Explicit vs. tacit
  collusion—the impact of communication in oligopoly experiments.}
  \emph{European Economic Review}, 56, 1759--1772.
\bibAnnoteFile{fonseca2012explicit}

\bibitem[{Fr{\'e}chette et~al.(2020)Fr{\'e}chette, Lizzeri, and
  Vespa}]{frechette2020multimarket}
Fr{\'e}chette, Guillaume, Alessandro Lizzeri, and Emanuel Vespa (2020),
  \enquote{Collusion in multimarkets: An experiment.} NYU Working Paper.
\bibAnnoteFile{frechette2020multimarket}

\bibitem[{Fudenberg et~al.(2010)Fudenberg, Rand, and
  Dreber}]{fudenberg2010slow}
Fudenberg, D., D.G. Rand, and A.~Dreber (2010), \enquote{Slow to anger and fast
  to forgive: Cooperation in an uncertain world.} \emph{American Economic
  Review}.
\bibAnnoteFile{fudenberg2010slow}

\bibitem[{Ghidoni and Suetens(forthcoming)}]{ghidoni2020effect}
Ghidoni, Riccardo and Sigrid Suetens (forthcoming), \enquote{The effect of
  sequentiality on cooperation in repeated games.} \emph{American Economic
  Journal: Microeconomics}.
\bibAnnoteFile{ghidoni2020effect}

\bibitem[{Goette and Schmutzler(2009)}]{goette2009merger}
Goette, Lorenz and Armin Schmutzler (2009), \enquote{Merger policy: What can we
  learn from competition policy.} \emph{Experiments and Competition Policy;
  Hinloopen, J., Normann, HT, Eds}, 185--216.
\bibAnnoteFile{goette2009merger}

\bibitem[{Harrington et~al.(2013)Harrington, Gonzalez, and
  Kujal}]{harrington2013relative}
Harrington, Joseph~E, Roberto~Hernan Gonzalez, and Praveen Kujal (2013),
  \enquote{The relative efficacy of price announcements and express
  communication for collusion: Experimental findings.} \emph{Working paper.
  University of Pennsylvania, The Wharton School}.
\bibAnnoteFile{harrington2013relative}

\bibitem[{Harrington et~al.(2016)Harrington, Gonzalez, and
  Kujal}]{harrington2016relative}
Harrington, Joseph~E, Roberto~Hernan Gonzalez, and Praveen Kujal (2016),
  \enquote{The relative efficacy of price announcements and express
  communication for collusion: Experimental findings.} \emph{Journal of
  Economic Behavior \& Organization}, 128, 251--264.
\bibAnnoteFile{harrington2016relative}

\bibitem[{Harsanyi and Selten(1988)}]{harsanyi1988general}
Harsanyi, John~C and Reinhard Selten (1988), \emph{A general theory of
  equilibrium selection in games}. MIT Press, Cambridge, MA.
\bibAnnoteFile{harsanyi1988general}

\bibitem[{Horstmann et~al.(2018)Horstmann, Kr{\"a}mer, and
  Schnurr}]{horstmann2018number}
Horstmann, Niklas, Jan Kr{\"a}mer, and Daniel Schnurr (2018), \enquote{Number
  effects and tacit collusion in experimental oligopolies.} \emph{Journal of
  Industrial Economics}, 66, 650--700.
\bibAnnoteFile{horstmann2018number}

\bibitem[{Huck et~al.(2007)Huck, Konrad, M{\"u}ller, and
  Normann}]{huck2007merger}
Huck, Steffen, Kai~A. Konrad, Wieland M{\"u}ller, and Hans-Theo Normann (2007),
  \enquote{The merger paradox and why aspiration levels let it fail in the
  laboratory.} \emph{Economic Journal}, 117, 1073--1095.
\bibAnnoteFile{huck2007merger}

\bibitem[{Huck et~al.(2004)Huck, Normann, and Oechssler}]{huck2004two}
Huck, Steffen, Hans-Theo Normann, and J{\"o}rg Oechssler (2004), \enquote{Two
  are few and four are many: number effects in experimental oligopolies.}
  \emph{Journal of Economic Behavior \& Organization}, 53, 435--446.
\bibAnnoteFile{huck2004two}

\bibitem[{Kartal and M{\"u}ller(2018)}]{kartal2018new}
Kartal, Melis and Wieland M{\"u}ller (2018), \enquote{A new approach to the
  analysis of cooperation under the shadow of the future: Theory and
  experimental evidence.} University of Vienna working paper.
\bibAnnoteFile{kartal2018new}

\bibitem[{Kartal et~al.(2017)Kartal, M{\"u}ller, and
  Tremewan}]{kartal2017building}
Kartal, Melis, Wieland M{\"u}ller, and James Tremewan (2017), \enquote{Building
  trust: The costs and benefits of gradualism.} University of Vienna working
  paper.
\bibAnnoteFile{kartal2017building}

\bibitem[{Kloosterman(2019)}]{kloosterman2019cooperation}
Kloosterman, Andrew (2019), \enquote{Cooperation in stochastic games: a
  prisoner’s dilemma experiment.} \emph{Experimental Economics}, 1--21.
\bibAnnoteFile{kloosterman2019cooperation}

\bibitem[{Lugovskyy et~al.(2017)Lugovskyy, Puzzello, Sorensen, Walker, and
  Williams}]{lugovskyy2017experimental}
Lugovskyy, Volodymyr, Daniela Puzzello, Andrea Sorensen, James Walker, and
  Arlington Williams (2017), \enquote{An experimental study of finitely and
  infinitely repeated linear public goods games.} \emph{Games \& Economic
  Behavior}, 102, 286--302.
\bibAnnoteFile{lugovskyy2017experimental}

\bibitem[{Marshall and Marx(2012)}]{marshall2012economics}
Marshall, Robert~C and Leslie~M Marx (2012), \emph{The economics of collusion:
  Cartels and bidding rings}. MIT Press.
\bibAnnoteFile{marshall2012economics}

\bibitem[{Perry and Porter(1985)}]{perry1985oligopoly}
Perry, Martin~K and Robert~H Porter (1985), \enquote{Oligopoly and the
  incentive for horizontal merger.} \emph{American Economic Review}, 75,
  219--227.
\bibAnnoteFile{perry1985oligopoly}

\bibitem[{Potters and Suetens(2013)}]{potters2013oligopoly}
Potters, Jan and Sigrid Suetens (2013), \enquote{Oligopoly experiments in the
  current millennium.} \emph{Journal of Economic Surveys}, 27, 439--460.
\bibAnnoteFile{potters2013oligopoly}

\bibitem[{Rosokha and Wei(2020)}]{rosokha2020cooperation}
Rosokha, Yaroslav and Chen Wei (2020), \enquote{Cooperation in queueing
  systems.} \emph{Working Paper}.
\bibAnnoteFile{rosokha2020cooperation}

\bibitem[{Salz and Vespa(2020)}]{salz2020estimating}
Salz, Tobias and Emanuel Vespa (2020), \enquote{Estimating dynamic games of
  oligopolistic competition: An experimental investigation.} \emph{RAND Journal
  of Economics}, 51, 447--469.
\bibAnnoteFile{salz2020estimating}

\bibitem[{Sherstyuk et~al.(2013)Sherstyuk, Tarui, and
  Saijo}]{sherstyuk2013payment}
Sherstyuk, Katerina, Nori Tarui, and Tatsuyoshi Saijo (2013), \enquote{Payment
  schemes in infinite-horizon experimental games.} \emph{Experimental
  Economics}, 16, 125--153.
\bibAnnoteFile{sherstyuk2013payment}

\bibitem[{Vespa(2020)}]{vespa2020experimental}
Vespa, Emanuel (2020), \enquote{An experimental investigation of cooperation in
  the dynamic common pool game.} \emph{International Economic Review}, 61,
  417--440.
\bibAnnoteFile{vespa2020experimental}

\bibitem[{Vespa and Wilson(2019)}]{vespa2019experimenting}
Vespa, Emanuel and Alistair~J Wilson (2019), \enquote{Experimenting with the
  transition rule in dynamic games.} \emph{Quantitative Economics}, 10,
  1825--1849.
\bibAnnoteFile{vespa2019experimenting}

\bibitem[{Vespa and Wilson(2020)}]{vespa2020experimenting}
Vespa, Emanuel and Alistair~J Wilson (2020), \enquote{Experimenting with
  equilibrium selection in dynamic games.} \emph{Working Paper}.
\bibAnnoteFile{vespa2020experimenting}

\bibitem[{Vesterlund(2016)}]{vesterlund2016using}
Vesterlund, Lise (2016), \enquote{Using experimental methods to understand why
  and how we give to charity.} \emph{Handbook of Experimental Economics}, 2,
  91--151.
\bibAnnoteFile{vesterlund2016using}

\bibitem[{Weber(2006)}]{weber2006managing}
Weber, Roberto~A (2006), \enquote{Managing growth to achieve efficient
  coordination in large groups.} \emph{American Economic Review}, 96, 114--126.
\bibAnnoteFile{weber2006managing}

\bibitem[{Wilson and Vespa(2020)}]{wilson2020information}
Wilson, Alistair~J. and Emanuel Vespa (2020), \enquote{Information transmission
  under the shadow of the future: An experiment.} \emph{American Economic
  Journal: Microeconomics}, 12.
\bibAnnoteFile{wilson2020information}

\end{thebibliography}

\appendix
\let\oldthetable\thetable
\let\oldthefigure\thefigure
\setcounter{table}{0}
\setcounter{figure}{0}
\renewcommand{\thetable}{A.\oldthetable}
\renewcommand{\thefigure}{A.\oldthefigure}
\vfill\eject
\section{Additional Tables and Figures\label{sec:app:tabs}}
\begin{table}[h]
    \caption{Cooperation rates (All supergames)}
    \centering
     \begin{tabular}{cccccc}\toprule
\multirow[c]{2}{*}{\minitab[c]{\textbf{Cooperation} \\ \textbf{rates}}} & \multicolumn{2}{c}{$x=\$9$} & &\multicolumn{2}{c}{$x=\$1$}\\ 
\cmidrule{2-3}\cmidrule{5-6}
 & $N=2$ & $N=4$ & & $N=4$ & $N=10$ \\ \midrule
  Initial coop. & $\mEst{0.466}{0.046}$ & $\mEst{0.100}{0.021}$ & & $\mEst{0.719}{0.039}$ & $\mEst{0.457}{0.044}$ \\ 
 Ongoing coop.& $\mEst{0.296}{0.029}$ & $\mEst{0.044}{0.012}$ & & $\mEst{0.433}{0.034}$ & $\mEst{0.243}{0.039}$
 \\ \cmidrule{2-6}
  Initial success & 0.466 & 0.003 & & 0.408 & 0.010 \\ 
 Ongoing success  & 0.296 & 0.002 & & 0.275 & 0.009  \\  
 \bottomrule 
\end{tabular}
\begin{tablenotes}
Results are calculated using data from all supergames. Cooperation rates present raw proportions (with subject-clustered standard errors). %The cooperation decomposition runs two subject-clustered probits on the cooperation decision (initial and ongoing) where variables are dummies for a low correlated basin treatment ($x=\$1$, both $N$ values) and a high independent basin treatment ($x=\$9/N=4$ and $x=\$1/N=10$). Coefficients shown are the predicted level at just the constant (the $p^\star_0$ column) and the predicted cooperation change in each RHS variable.
\end{tablenotes}

    \label{tab:CoopAllSG}
\end{table}
\begin{table}[h]
    \caption{Cooperation in reaction to previous round's history}
    \centering
     \begin{tabular}{ccccccccc}\toprule
 & \multicolumn{2}{c}{$x=\$9$} & &\multicolumn{2}{c}{$x=\$1$} & &\multicolumn{2}{c}{Chat ($x=\$9,N=4$)}\\ 
\cmidrule{2-3}\cmidrule{5-6}\cmidrule{8-9}
 History & $N=2$ & $N=4$ & & $N=4$ & $N=10$ & & $\delta=\nicefrac{3}{4}$ & $\delta=\nicefrac{1}{2}$\\ \midrule
$(C,S)$ & $\mEst{0.977}{0.011}$ & -- & &$\mEst{0.988}{0.013}$ & -- & &$\mEst{0.980}{0.006}$ & $\mEst{0.750}{0.217}$\\ 
$(C,F)$ & $\mEst{0.317}{0.063}$ & $0.000$ & &$\mEst{0.521}{0.085}$ & $\mEst{0.739}{0.077}$ & &$\mEst{0.342}{0.0073}$ & $\mEst{0.255}{0.104}$ \\
$(D,S)$ & $\mEst{0.150}{0.060}$ & -- & & $\mEst{0.263}{0.110}$ & -- & &$\mEst{0.143}{0.136}$ & $\mEst{0.750}{0.217}$\\
$(D,F)$  & $\mEst{0.033}{0.006}$ & $\mEst{0.006}{0.004}$ & &$\mEst{0.023}{0.009}$ & $\mEst{0.025}{0.009}$ & &$\mEst{0.019}{0.019}$ & $\mEst{0.006}{0.004}$ \\
\bottomrule
\end{tabular}
\begin{tablenotes}
Data taken from last five supergames in each treatment (with subject-clustered standard errors). Cells marked ``--'' have no observations at the relevant history. History shows the own-action-signal pair from the previous round, $(a_{t-1},\sigma_{t-1})$.
\end{tablenotes}
    \label{tab:history}
\end{table}

\begin{figure}[tb]
    \centering
    \subfloat[Initial cooperation]{
    \includegraphics[width=0.49\textwidth]{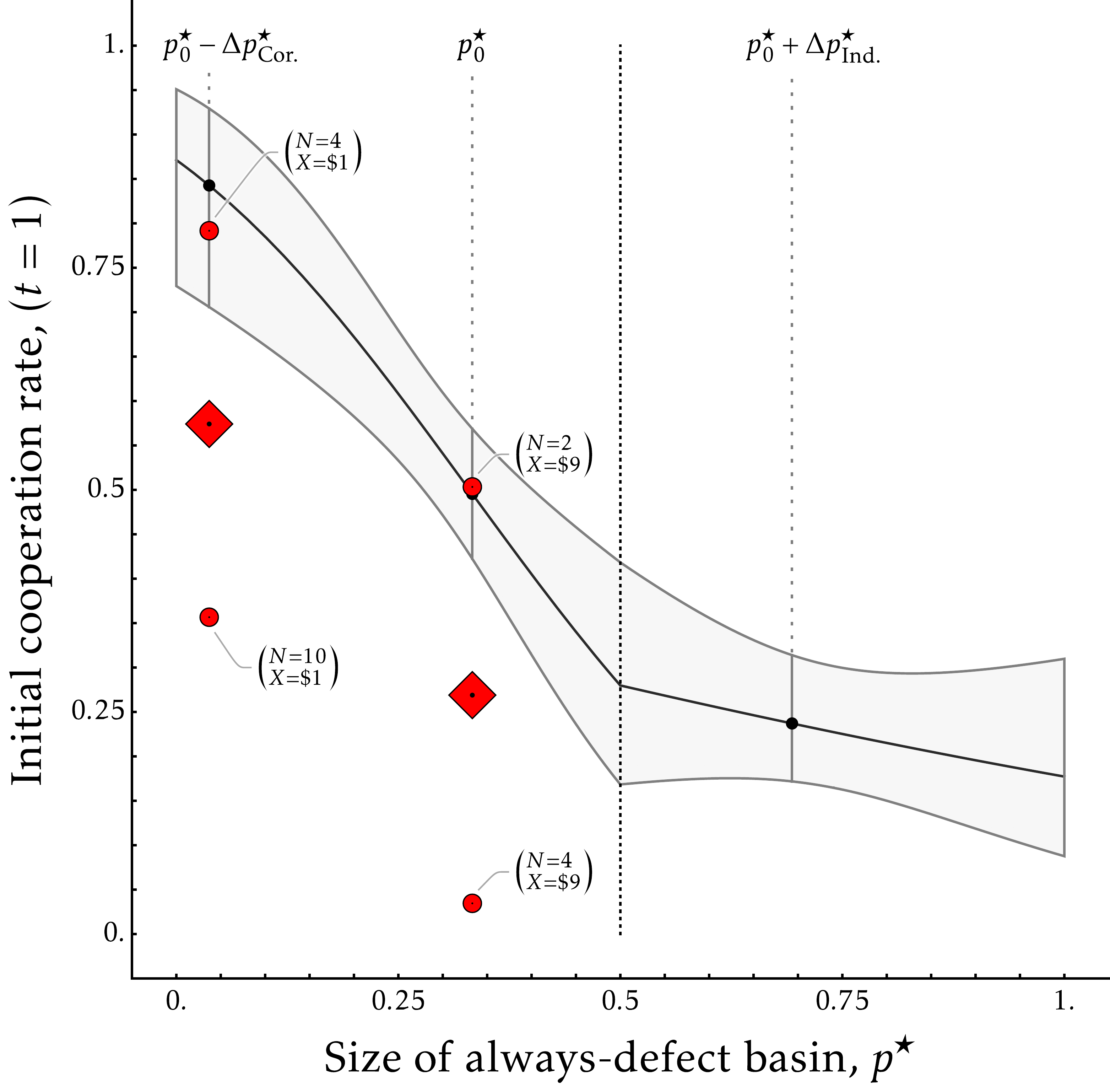}}
    \subfloat[Ongoing cooperation]{
    \includegraphics[width=0.49\textwidth]{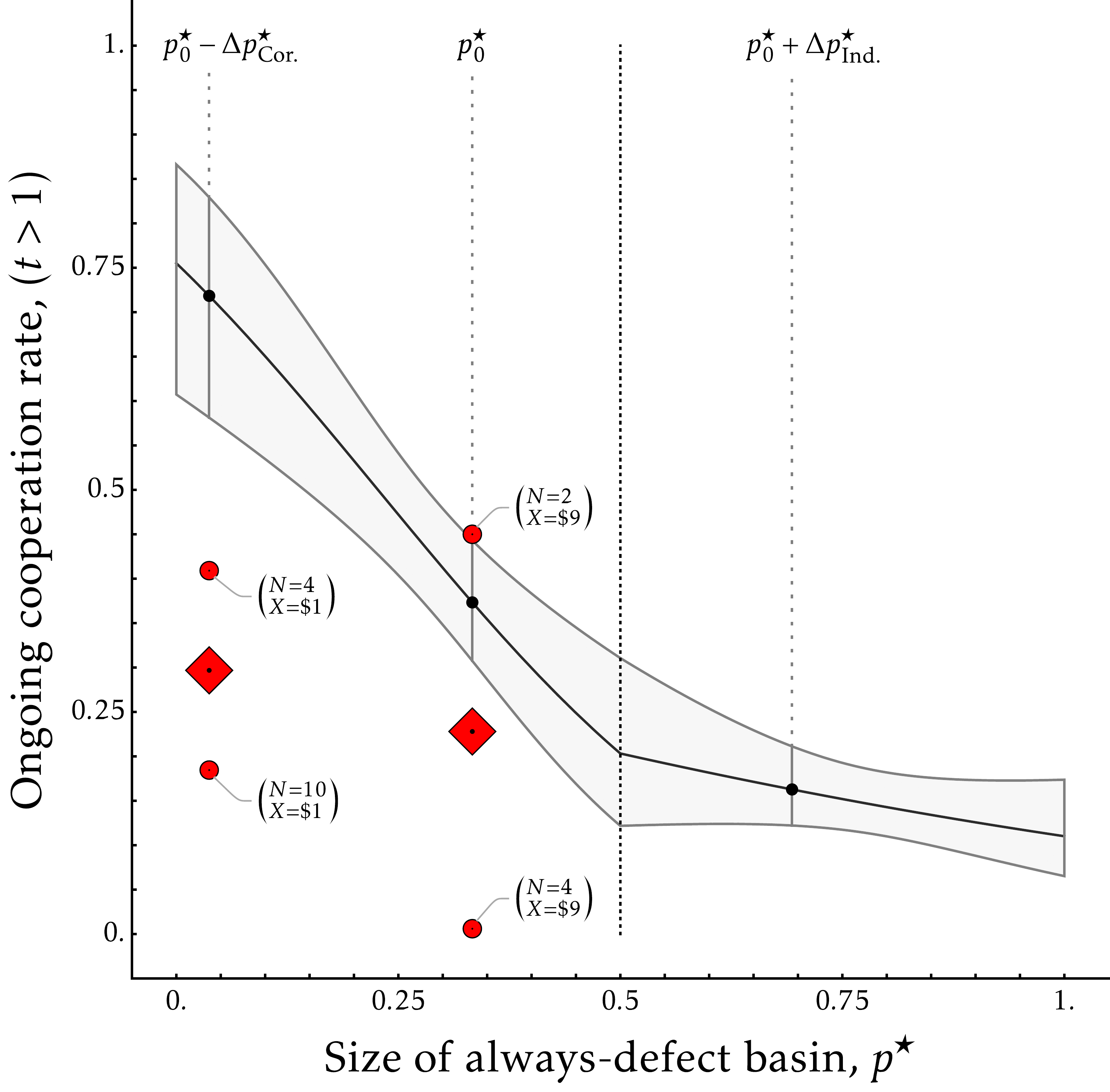}}
    \caption{Cooperation and the Correlated Basin-Size Model}
    \label{fig:ResultsCorr}
    
\begin{tablenotes}
Figures show pooled data by the correlated basin (diamonds) while the separate treatments are illustrated as the surrounding circles. See Figure \ref{fig:ResultsInd} in the main paper for the analogous figure under the independent basin.
\end{tablenotes}
\end{figure}

\pagebreak
\section{Further Analysis of the Within-Subject Treatments}
We do find evidence that is hysteresis in the dynamics though. A large and immediate jump in cooperation is observed as the game moves from $N=4$ to $N=2$, but with no initial response when $N$ moves in the opposite direction. This suggests that in the short run, absent experience with the new environment subjects are likely to try to coordinate on cooperation. Notice, however, that the response to the change in $N$ in the initial supergames of the second half can be compared against the initial supergames of the first half. Shifting the illustrated behavior for supergames 11--20 in Figure \ref{fig:Extensions}(A) ten supergames to the left and comparing the observed levels to supergames 1--10 makes the jump patterns less clear. In the first ten supergames under $N=4$ the pattern is very similar: cooperation is initially high, but it falls rapidly as the subjects gain experience. Similarly for the $N=2$ trend. 

While some caution is warranted as the qualitative trends as they gain experience under the new parameterization are similar to the initial trends with no prior experience,  direct comparisons of behavior in supergames one and eleven do reject equivalence. With no experience at all in the environment, 43.2 percent of subjects cooperate for $N=2$ in the first round of the first supergame, compared to 29.4 percent for $N=4$ (significantly different with $p=0.005$ from a test of proportions). In contrast to the significant difference over $N$ in the very first decision, in supergame 11 of our within-subject sessions (with prior experience at an alternate value of $N$) the  initial cooperation rates at $N=2$ and $N=4$ can not be distinguished from one another in the plotted figure (at 60.0 and 59.7 percent cooperation, respectively), let alone statistically ($p=0.974$). Both cooperation rates are significantly greater than the initial responses in supergame one.\footnote{Given the disjoint subject groups and identical treatment in supergames 1--10, we compare proportions using $t$-tests without clustering. We then compare the initial response under each value of $N$ in the within-subject supergame eleven to all subjects at that $N$ in supergame one. Using these tests we reject equivalence with $p=0.021$ for $N=2$ and $p<0.001$ for $N=4$.} We conclude that experience at another parameter value in the first half does cause both treatments' cooperation rates to increase.

\begin{table}[tb!]
    \centering
    \caption{Cooperation: Between vs. Within}
    \begin{tabular}{ccccccccc}
\toprule
 & \multicolumn{2}{c}{Between (SG 6--10)} & &\multicolumn{2}{c}{Within (SG 16--20)} & $\Delta_{\text{Btwn.}}$ &  \multicolumn{2}{c}{$\Delta_{\text{Wthn.}}$}\\ 
\cmidrule{2-3}\cmidrule{5-6}\cmidrule{8-9}
 & $N=2$ & $N=4$ & & $N=2$ & $N=4$ & & $2\rightarrow 4$ & $4\rightarrow 2$  \\ \midrule
  Initial coop. & $\mEst{0.474}{0.036}$ & $\mEst{0.139}{0.025}$ & & $\mEst{0.643}{0.056}$ & $\mEst{0.214}{0.041}$ 
  &$\mEst{-0.469}{0.060}$ & $\mEst{-0.260}{0.042}$ & $\mEst{-0.504}{0.056}$ \\ 
 Ongoing coop. & $\mEst{0.299}{0.026}$ & $\mEst{0.054}{0.012}$ & & $\mEst{0.598}{0.051}$ & $\mEst{0.042}{0.016}$
  & $\mEst{-0.444}{0.055}$ & $\mEst{-0.258}{0.029}$ & $\mEst{-0.544}{0.050}$  
\\ \cmidrule{2-9}
 Initial success & 0.474 & 0.011 &  & 0.643 & 0.042  
 & -0.503 & -0.433 & -0.632  \\ 
 Ongoing success  & 0.299 & 0.004 &  & 0.598 & 0.008
 & -0.450 & -0.292 & -0.594   \\  
 \bottomrule
\end{tabular}
\begin{tablenotes}
Comparisons at the same experience level are generated using supergames 6--10 across all sessions (fixing $N$, between and within sessions are identical until supergame 11). For the within change we measure the cooperation rates in supergames 16--20. All cooperation rates are raw proportions (with subject-clustered standard errors). The last three columns measure the corresponding cooperation rate when $N=4$ minus the cooperation rate when $N=2$. 
\end{tablenotes}

%  & \multicolumn{3}{c}{Between (SG 6--10)} & &\multicolumn{3}{c}{Within (SG 16--20)}\\ 
% \cmidrule{2-4}\cmidrule{6-8}
%  & $N=2$ & $N=4$ & $\Delta$ & & $N=2$ & $N=4$ & $\Delta$ \\ \midrule
%   Initial coop. & $\mEst{0.474}{0.036}$ & $\mEst{0.139}{0.025}$ &$\mEst{-0.335}{0.044}$ & & $\mEst{0.643}{0.056}$ & $\mEst{0.214}{0.041}$ & $\mEst{-0.429}{0.069}$\\ 
%  Ongoing coop. & $\mEst{0.299}{0.026}$ & $\mEst{0.054}{0.012}$ &$\mEst{-0.246}{0.029}$ & & $\mEst{0.598}{0.051}$ & $\mEst{0.042}{0.016}$& $\mEst{-0.556}{0.053}$
% \\ \cmidrule{2-8}
%  Initial success & 0.474 & 0.011 & -0.463 & & 0.643 & 0.042  & -0.601 \\ 
%  Ongoing success  & 0.299 & 0.004 & -0.296 & & 0.598 & 0.008 & -0.590   \\  
%  \bottomrule
% \end{tabular}
    \label{tab:Within}
\end{table}

In Table \ref{tab:Within} we provide more details, where we compare and contrast the behavior after 5 rounds of experience. In the first two data columns we present average behavior (initial/ongoing cooperation and success, with subject-clustered standard errors for the individual choices) in supergames 6--10 for $N=2$, $N=4$. In the next column pair we present the same information in supergames 16--20 for the within treatments only. Examining the differences across the \emph{within} and \emph{between} cooperation levels, while we find no significant differences in behavior for $N=4$ ($p=0.117/p=0.539$ for initial/ongoing cooperation) we do find significant differences across the $N=2$ cooperation rates ($p=0.011$ for initial, $p<0.001$ for ongoing). The significant differences here reflect the substantially greater upward shift in the $4\rightarrow 2$ treatment.

In the final three columns, we compute (for three different cases) the average cooperation rate in supergames when $N=4$ minus the cooperation rate when $N=2$. In the first column ($\Delta_{\text{Btwn}}$) we calculate the between-subject change using data from supergames 16--20 in the $X=\$9$ between-subject treatments. The results here are analogous to the marginal effect attributable to an increase in the independent basin of $\Delta p^\star_{\text{Ind.}}=+0.36$ in Table \ref{tab:Aggregate} once we remove the $X=\$1$ treatments. In the final column pair we present the same assessed treatment effect if we used the within-subject difference in the $2\rightarrow 4$  and $4\rightarrow 2$ treatments (here we compare data in supergames 6--10 to supergames 16--20).

While the three measures agree qualitatively---and exhibit economically large effects in $N$ in the same direction---there are differences, particularly in the comparisons to the $2\rightarrow 4$ case. However, we note that there are two effects at play here. In the $2\rightarrow 4$ comparison, reduced magnitudes are driven primarily by the fact that behavior in this treatment has not converged. To see this, consider the assessed between-subject effect if we used data from supergames 6--10: a -33.5 percentage point effect on initial cooperation, which is not significantly different from the -26.0 percent effect identified in the within comparison ($p=0.117$).\footnote{Similarly for ongoing cooperation the between-effect assessed in supergames 6--10 is -24.6 percent compared to -25.8 percent within ($p=0.539$).} In contrast, the greater assessed effect in the $4\rightarrow 2$ comparison is the composite of the same \emph{reduction} in the effect from looking at the still-converging data for $N=4$, with a substantial increase in cooperation at $N=2$ in the second half over the between-subject levels.

\begin{figure}[tb!]
\centering
    \subfloat[Between vs. Within ($x=\$9$)]{
    \includegraphics[width=0.49\textwidth]{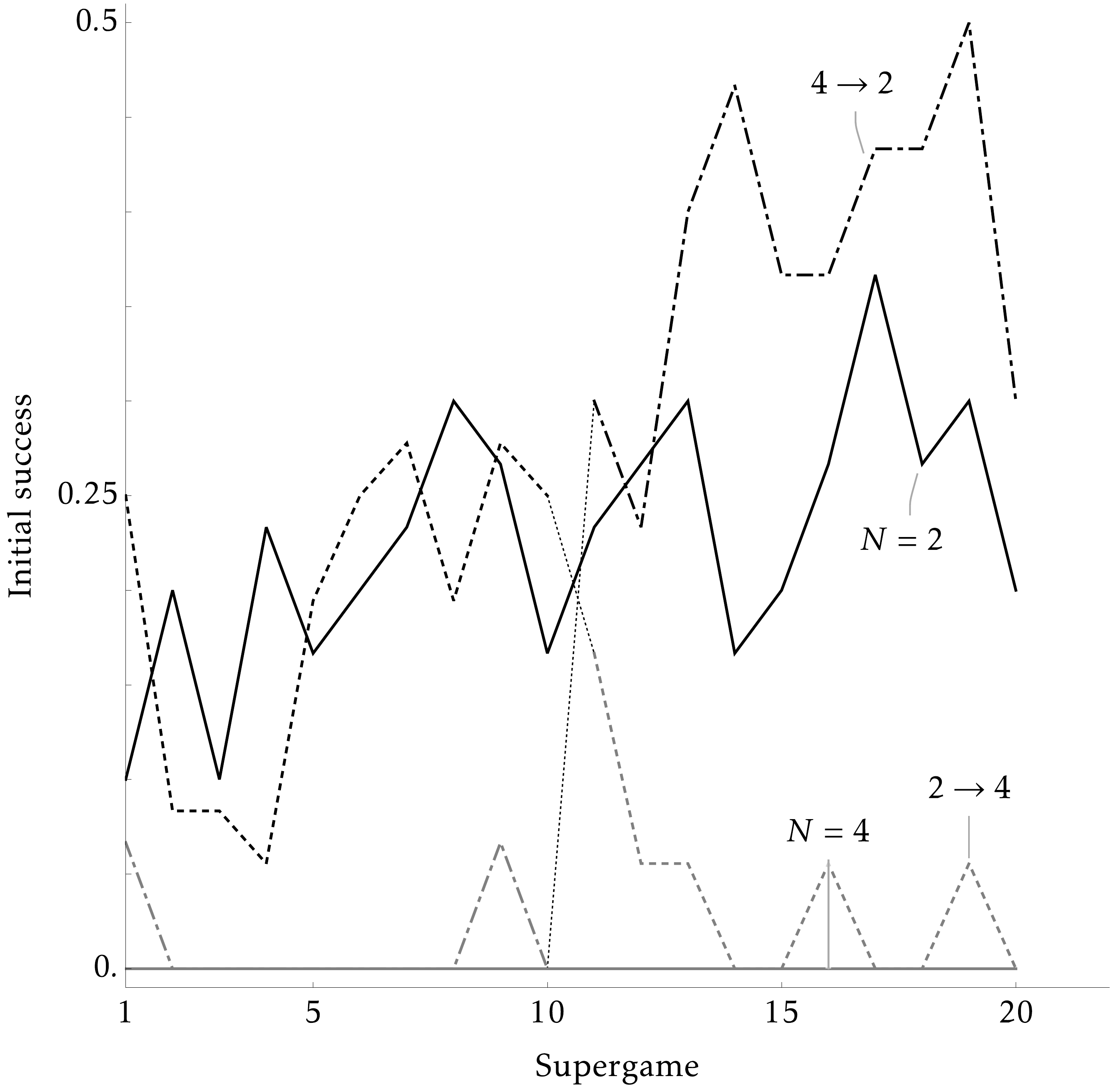}}
    \subfloat[Explicit vs. Implicit ($N=4; x=\$9$)]{
    \includegraphics[width=0.49\textwidth]{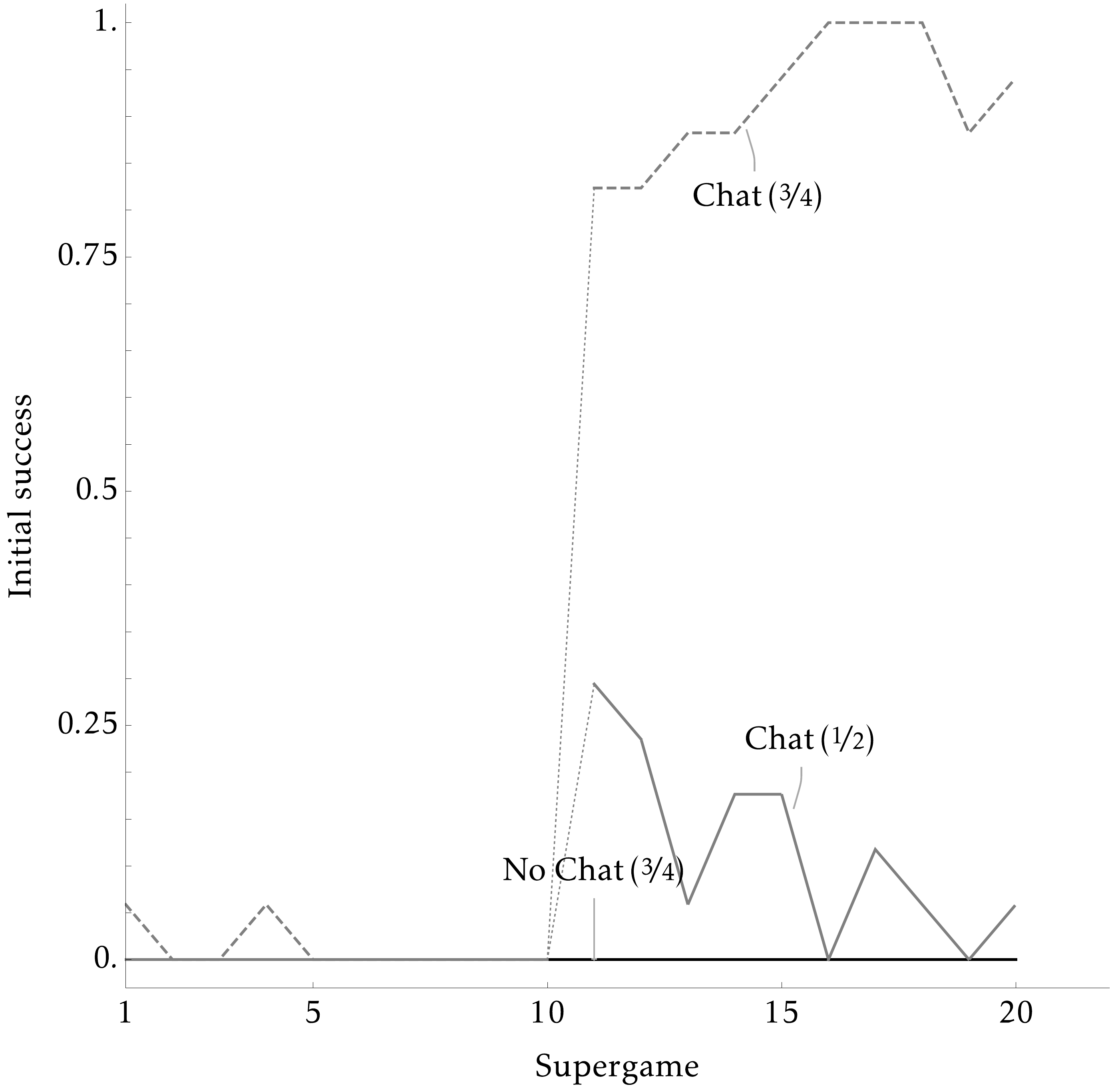}}
    \caption{Initial success rates in extensions (by supergame)}
    \label{fig:ExtensionsApp}
\end{figure}

\pagebreak
\section{Interface Screenshots}
\begin{figure}[h]
    \centering
    \subfloat[Action Selection]{
    \includegraphics[width=0.75\textwidth]{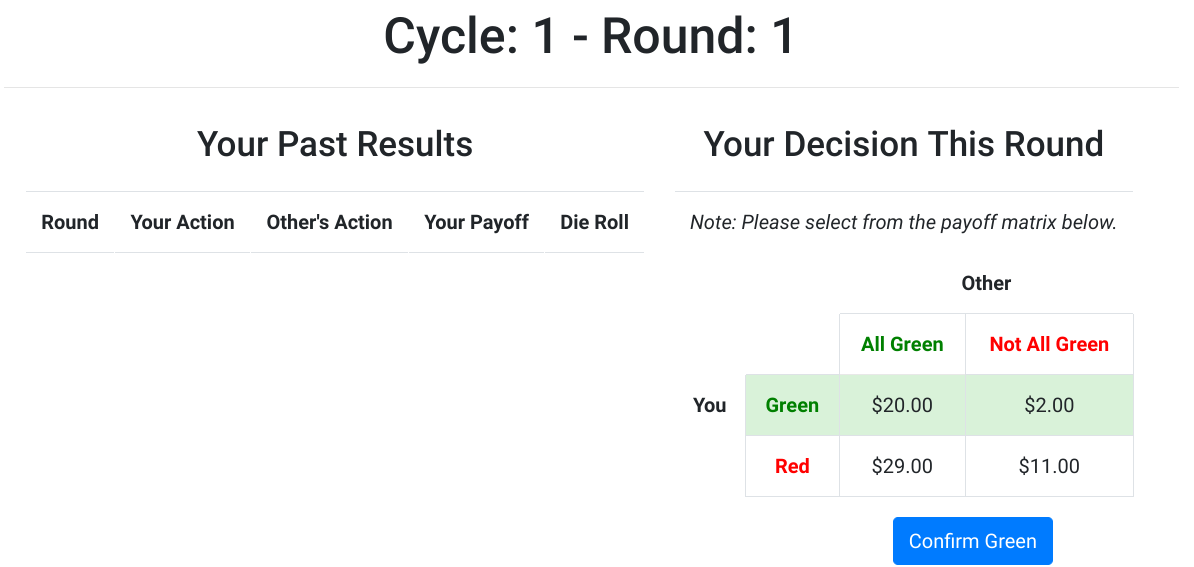}}
    
    \subfloat[Round Feedback]{
    \includegraphics[width=0.75\textwidth]{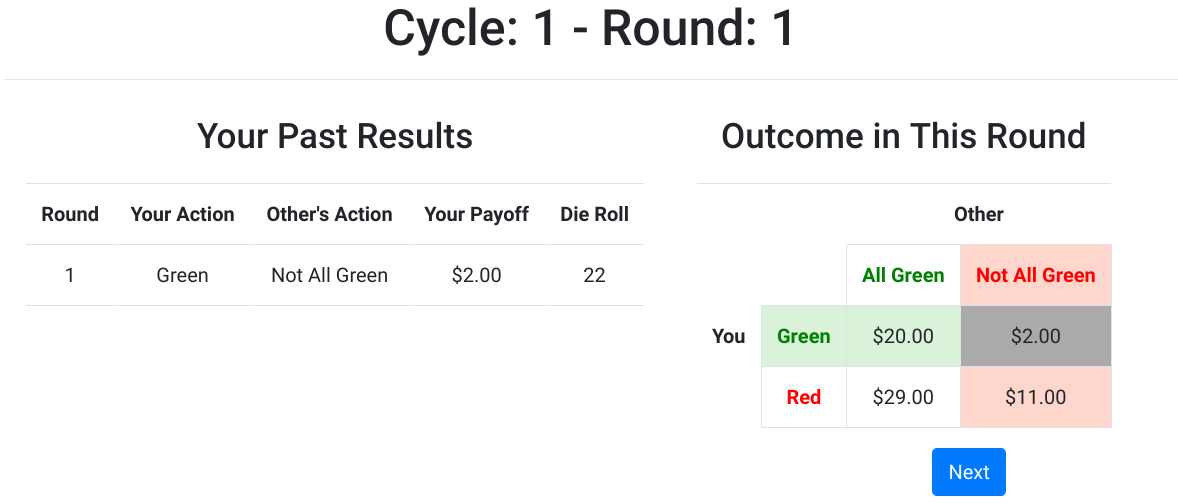}}
    
    \subfloat[Supergame Feedback]{
    \includegraphics[width=0.75\textwidth]{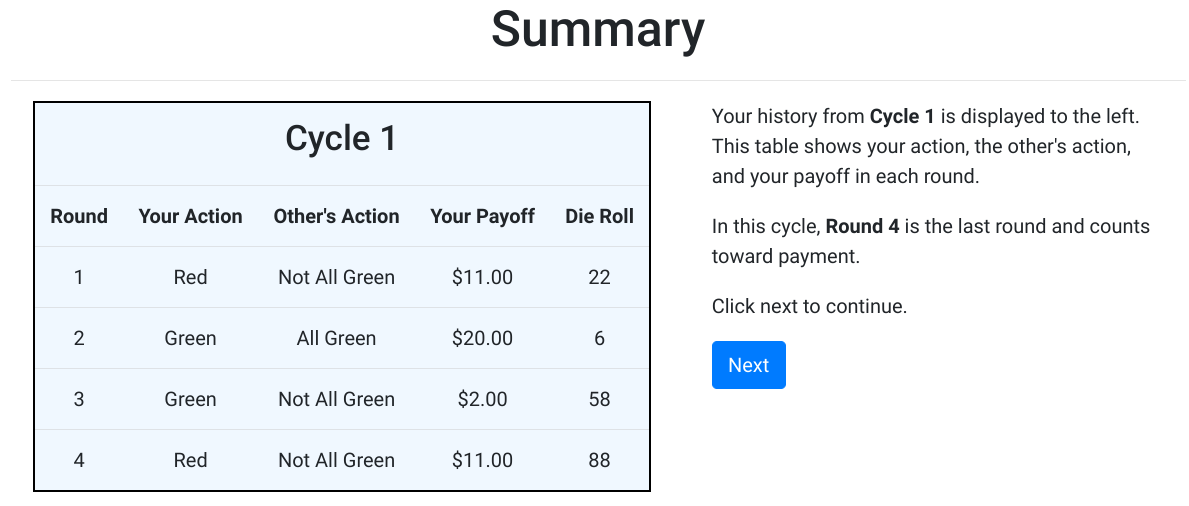}}
    \caption{Interface Screenshots}
    \label{fig:Screenshots}
\end{figure}
\pagebreak
\singlespacing

\section{Provided Instructions\label{sec:app:instructions}}
Below we include the instructions given to participants. All language deltas/treatment-specific language are included in braces. Text in red is for the $N=2$ treatment, in blue for the $N>2$ treatments (here we provide the $N=4$ implementation, where $N=10$ has only minor changes.). In green we provide the payoff text for $X=\$9$, in orange for $X=\$1$. Separate instructions for \{Part two\} are given by treatment for the extensions. The only unlisted treatment variation here is for the Chat($\nicefrac{1}{2}$) treatment, where the only changes are for the critical die rolls in the \emph{Study Organization \& Payment} section, where the supergame cutoff changes from 75 to 50.
\section*{Instructions}
\subsection*{Welcome}
You are about to participate in a study on decision-making. What you earn depends on your decisions, and the decisions of others in this room. Please turn off your cell phones and any similar devices now. Please do not talk or in any way try to communicate with other participants. We will start with a brief instruction period. During the instruction period you will be given a description of the main features of the study. If you have any questions during this period, raise your hand and your question will be answered in private at your computer carrel.

\subsection*{Study Organization \& Payment}
\begin{itemize}
	\item The study has two Parts, where each Part has 10 decision-making \textbf{Cycles}. Each Cycle consists of a random number of \textbf{Rounds} where you make decisions.
	\item At the end of the study, one of the two Parts will be selected for payment with equal probability. For the selected Part, one of the 10 Cycles will be randomly selected for payment. Your payment for this randomly selected Cycle will be based on your decision's in that Cycle's last Round.
	
	\item The number of Rounds in each Cycle is random, where only the last Round in each Cycle counts for payment. Which Round is the last is determined as follows:
    \begin{itemize}
        \item In every Round, after participants make their decisions, the computer will roll a fair 100-sided die. If the die roll is greater than\terminalText{50}{75} (so\terminalText{51}{76}--100) the round just completed is the one that is used to determine the current Cycle's payment, and the Cycle ends. If instead the computer's roll is less than\terminalText{50}{75} (so 1--\terminalText{50}{75}) then the Cycle continues into another Round. 
		\item Because of this rule, after every Round decision there is a\terminalText{50}{25} percent chance that the current Round is the ones that count for the Cycle's payment, and a\terminalText{50}{75} percent chance that the Cycle continues and the decisions in a subsequent round will count for that Cycle payment.
    \end{itemize}
	\item Your final payment for the study will be made up of a \$6 show-up fee, and your payment from the last Round in the randomly selected Cycle.
\end{itemize}

\subsection*{Part 1}
\begin{itemize}
    \item In the first part of the study you will make decisions in 10 Cycles. In each Cycle you will be matched with\groupText{another participant}{a group of three other participants}{a group of nine other participants}in the room for a sequence of Rounds. You will interact with the same\groupText{other participant}{group of three other participants}{group of nine other participants}in all rounds of the cycle.
    \item Once a Cycle is completed, you will be randomly matched to a new\groupText{participant}{group of three participants}{group of nine participants}for the next Cycle. 
	\item While the specific\groupText{participant}{participants}{participants}you are matched to is fixed across all Rounds in the Cycle, the computer interface in which you make your decisions is anonymous, so you will never find out which participants in the room you interacted with in a particular Cycle, nor will others be able to find out that they interacted with you.
\end{itemize}

\subsection*{Round Choices and Payoffs}
For each Round in each Cycle, you and the matched\groupText{participant}{participants in your group}{participants in your group}will make simultaneous choices.\groupText{Both}{All four}{All ten}of you must choose between either the \textbf{Green} action or the \textbf{Red} action. After you and the other\groupText{participant}{three participants}{nine participants}have made your choices, you will be given feedback on the\groupText{other participant's}{other participants'}{other participants'}choices that Round, alongside the Computer's die roll to determine if that Round counts for the Cycle payment. 

If a particular Round is the Cycle's last, and that Cycle is the one selected for final payment, there are four possible payoff outcomes.

\begin{enumerate}
	\item If both you and\groupText{the other participant}{\antibText{all three}{any} of the other participants}{\antibText{all nine}{any} of the other participants}choose the Green action, you get a round payoff of \$20.
	\item If you choose the Green action and\groupText{the other participant chooses}{\antibText{any}{all} of the other participants choose}{\antibText{any}{all} of the other participants choose}Red, you get a round payoff of \payText{\$2}{\$10}.
	\item If you choose the Red action and\groupText{the other participant chooses}{\antibText{all}{any} of the three other participants choose}{\antibText{all}{any} of the nine other participants choose}Green, you get a round payoff of \payText{\$29}{\$21}.
	\item If both you and\groupText{the other participant}{\antibText{any}{all} of the other three participants}{\antibText{any}{all}of the other nine participants}choose the Red action, you get a round payoff of \$11.
\end{enumerate}

These four payoffs are summarized in the following table:
\begin{center}
\centering
\begin{tabular}{p{1in}rp{2in}p{2in}}
 & &\multicolumn{2}{c}{Other\groupText{Participant's Action:}{Participants' Actions:}{Participants' Actions:}}\\

	    &  									&  \multicolumn{1}{c}{\groupText{Green}{\antibText{All}{Any of} 3 Green}{\antibText{All}{Any of} 9 Green}} 	&  \multicolumn{1}{c}{\groupText{Red}{\antibText{Any of}{All} 3 Red}{\antibText{Any of}{All} 9 Red}}  \\
		 \cline{3-4}
\multirow{2}{*}{Your Action:} 	&	 Green  &  \multicolumn{1}{|c|}{\$20} 				& \multicolumn{1}{c|}{ \payText{\$2}{\$10}}			\\
\cline{3-4}
								& Red    	&  \multicolumn{1}{|c|}{ \payText{\$29}{\$21}}			& \multicolumn{1}{c|}{\$11}			\\
\cline{3-4}
\end{tabular}
\end{center}
Some examples of these payoffs:

\textbf{Case 1.} Suppose you choose Green and\groupText{the other participant}{all three of the other participants}{all nine of the other participants}in the Cycle also choose Green. If that Round is the final one in the Cycle\groupText{both}{all four}{all ten}of you would get a payoff of \$20.

\textbf{Case 2.} Suppose\groupText{you}{you and two of the other participants}{you and eight of the other participants}choose Green while the other participant chooses Red. If that Round is the final one in the Cycle\groupText{you}{you and the other two participants who chose Green}{you and the other eight participants who chose Green} would get a payoff of \antibText{\payText{\$2}{\$10}}{\$2}, while the other participant would get a payoff of \payText{\$29}{\$21}.

\textbf{Case 3.} Suppose you choose\antibText{Red}{Green} while\groupText{the other participant chooses}{all three of the other participants choose}{all nine of the other participants choose}\antibText{Green}{Red}. If that Round is the final one in the Cycle you would get a payoff of \antibText{\payText{\$29}{\$21}}{\payText{\$2}{\$10}}, while the other\groupText{participant}{three participants}{nine participants}would get a payoff of \antibText{\payText{\$2}{\$10}}{\$29}.

\textbf{Case 4.} Suppose you and\antibText{\groupText{the other participant choose Red.}{another participant choose Red while the other two participants choose Green.}{another participant choose Red while the other eight participants choose Green.}}{\groupText{the other participant choose Red.}{the other participants choose Red.}{the other participants choose Red.}}If that Round is the final one in the Cycle\antibText{\groupText{you}{you and the other participant that chose Red}{you and the other participant that chose Red}would get a payoff of\groupText{\$11}{\$11, while the other two participants would get a payoff of\payText{\$2}{\$10}}{\$11, while the other eight participants would get a payoff of\payText{\$2}{\$10}}.}{\groupText{you}{you and the other participants}{you and the other participants}would get a payoff of \$11.}

\subsection*{Part 2}
After Part 1 is concluded, you will be given instructions on Part 2, which will have a very similar structure to the task in Part 1.

\{END OF PART 1 HANDOUT\}

\subsection*{Part 2 Instructions \{Between Only, handed out Supergame 11\}}
Part 2 is identical to Part 1\chatText{.}{except for the beginning of each cycle where we will now allow the matched participants to chat to one another before the cycle begins.} In each of the 10 Cycles in Part 2 you will again be matched to\groupText{another participant}{three other participants}{nine other participants}in the room.

Similar to Part 1, the Cycle payoff is determined by the last round in the Cycle, where the payoff depends on the action you chose and the\groupText{action chosen by the matched participant}{actions chosen by the three matched participants}{actions chosen by the nine matched participants}for that Cycle. Similar to Part 1, the below Table summarizes the payoff based upon the choices made in the Cycle's last round.
\begin{center}
\centering
\begin{tabular}{p{1in}rp{2in}p{2in}}
 & &\multicolumn{2}{c}{Other\groupText{Participant's Action:}{Participants' Actions:}{Participants' Actions:}}\\

	    &  									&  \multicolumn{1}{c}{\groupText{Green}{\antibText{All}{Any of} 3 Green}{\antibText{All}{Any of} 9 Green}} 	&  \multicolumn{1}{c}{\groupText{Red}{\antibText{Any of}{All} 3 Red}{\antibText{Any of}{All} 9 Red}}  \\
		 \cline{3-4}
\multirow{2}{*}{Your Action:} 	&	 Green  &  \multicolumn{1}{|c|}{\$20} 				& \multicolumn{1}{c|}{ \payText{\$2}{\$10}}			\\
\cline{3-4}
								& Red    	&  \multicolumn{1}{|c|}{ \payText{\$29}{\$21}}			& \multicolumn{1}{c|}{\$11}			\\
\cline{3-4}
\end{tabular}
\end{center}

\{END OF PART 2 HANDOUT\}

\subsection*{Part 2 Instructions \{Within Only, handed out Supergame 11\}}
Part 2 is very similar to Part 1. However, in each of the 10 Cycles in Part 2 you will instead be matched to three other participants in the room for each Cycle.

Similar to Part 1, the Cycle payoff is determined by the last round in the Cycle, where the payoff depends on the action you chose and the actions chosen by the three matched participants for that Cycle. If a particular Round is the Cycle’s last, and that Cycle is the one selected for final payment, there are four possible payoff outcomes.

\begin{enumerate}
	\item If both you and all three of the other participants choose the Green action, you get a round payoff of \$20.
	\item If you choose the Green action and any of the other participants chooses Red, you get a round payoff of \$2.
	\item If you choose the Red action and all three other participants choose Green, you get a round payoff of \$29.
	\item If both you and any of the other three participants choose the Red action, you get a round payoff of \$11.
\end{enumerate}

These four payoffs are summarized in the following table:
\begin{center}
\centering
\begin{tabular}{p{1in}rp{2in}p{2in}}
 & &\multicolumn{2}{c}{Other Participant's Action:}\\

	    &  									&  \multicolumn{1}{c}{All 3 Green} 	&  \multicolumn{1}{c}{Any of 3 Red}  \\
		 \cline{3-4}
\multirow{2}{*}{Your Action:} 	&	 Green  &  \multicolumn{1}{|c|}{\$20} 				& \multicolumn{1}{c|}{ \$2}			\\
\cline{3-4}
								& Red    	&  \multicolumn{1}{|c|}{ \$29}			& \multicolumn{1}{c|}{\$11}			\\
\cline{3-4}
\end{tabular}
\end{center}
Some examples of these payoffs:

\textbf{Case 1.} Suppose you choose Green and all three of the other participants in the Cycle also choose Green. If that Round is the final one in the Cycle all four of you would get a payoff of \$20.

\textbf{Case 2.} Suppose you and two of the other participants choose Green while the other participant chooses Red. If that Round is the final one in the Cycle you and the other two participants who chose Green would get a payoff of \$2, while the other participant would get a payoff of \$29.

\textbf{Case 3.} Suppose you choose Red while all three of the other participants choose Green. If that Round is the final one in the Cycle you would get a payoff of \$29, while the other three participants would get a Round payoff of \$2.

\textbf{Case 4.} Suppose you and another participant choose Red while the other two participants choose Green. If that Round is the final one in the Cycle you and the other participant that chose Red would get a payoff of \$11, while the other two participants would get a payoff of \$2.  

\{END OF PART 2 HANDOUT\}

\subsection*{Part 2 Instructions \{Chat Only, handed out Supergame 11\}}
Part 2 is identical to Part 1 except for the beginning of each cycle where we will now allow the matched participants to chat to one another before the cycle begins. In each of the 10 Cycles in Part 2 you will again be matched to three other participants in the room.

Similar to Part 1, the Cycle payoff is determined by the last round in the Cycle, where the payoff depends on the action you chose and the actions chosen by the three matched participants for that Cycle. Similar to Part 1, the below Table summarizes the payoff based upon the choices made in the Cycle's last round.
\begin{center}
\centering
\begin{tabular}{p{1in}rp{2in}p{2in}}
 & &\multicolumn{2}{c}{Other Participants' Actions:}\\

	    &  									&  \multicolumn{1}{c}{All 3 Green} 	&  \multicolumn{1}{c}{Any of 3 Red}  \\
		 \cline{3-4}
\multirow{2}{*}{Your Action:} 	&	 Green  &  \multicolumn{1}{|c|}{\$20} 				& \multicolumn{1}{c|}{ \$2}			\\
\cline{3-4}
								& Red    	&  \multicolumn{1}{|c|}{ \$29}			& \multicolumn{1}{c|}{\$11}			\\
\cline{3-4}
\end{tabular}
\end{center}

In contrast to Part 1 though, at the beginning of each new cycle, a chat window will be given to you, which will stay open for two minutes, or until all group members close it.

You may not use the chat to discuss details about your previous earnings, nor are you to provide any details that may help other participants in this room identify you. This is important to the validity of this study and will be not tolerated. However, you are encouraged to use the chat window to discuss the upcoming Cycle.

If at any point within the two-minute limit you wish to leave the chat, you can click the ``Finish Chat'' button. The other participants will be informed that you left.

\{END OF PART 2 HANDOUT\}

% NEED TO ADD SCREENSHOTS AND INSTRUCTIONS 

\end{document}